\documentclass[journal]{IEEEtran}

\usepackage{cite}

\usepackage{amsmath}
\usepackage{mathrsfs}

\usepackage{algorithmic}

\usepackage{array}

\usepackage{booktabs}

\usepackage{subfigure}

\ifCLASSOPTIONcompsoc
\usepackage[caption=false,font=normalsize,labelfont=sf,textfont=sf]{subfig}
\else
\usepackage[caption=false,font=footnotesize]{subfig}
\fi

\usepackage{url}
\usepackage{graphicx,amsmath,amssymb,amsfonts}
\usepackage{algorithmic,algorithm}

\usepackage{setspace}

\usepackage{xcolor}

\usepackage{multirow}

\usepackage{diagbox}

\usepackage{xcolor}
\usepackage{xpatch}

\newtheorem{theorem}{\textbf{Theorem}}

\newtheorem{lemma}{\textbf{Lemma}}

\newtheorem{corollary}{\textbf{Corollary}}

\newcommand{\tabincell}[2]{\begin{tabular}{@{}#1@{}}#2\end{tabular}}

\makeatletter

\newcommand{\Rmnum}[1]{\expandafter\@slowromancap\romannumeral #1@}
\makeatother

\usepackage{framed} 

\usepackage{cancel}

\begin{document}
\bstctlcite{ref:BSTcontrol}

\title{Convergence Analysis and Latency Minimization for Semi-Federated Learning in Massive IoT Networks}

\author{
		Jianyang~Ren,~Wanli~Ni,\textit{~Member,~IEEE},~Hui~Tian,\textit{~Senior~Member,~IEEE},~and~Gaofeng~Nie,\textit{~Member,~IEEE}
		
		\thanks{		
		This work was supported by the National Natural Science Foundation of China under Grant 62071068. (Corresponding author: Hui Tian.)
		
		J. Ren, W. Ni, H. Tian and G. Nie are with the State Key Laboratory of Networking and Switching Technology, Beijing University of Posts and Telecommunications,	Beijing 100876, China (e-mail: renjianyang@bupt.edu.cn; charleswall@bupt.edu.cn; tianhui@bupt.edu.cn; niegaofeng@bupt.edu.cn).}
		}

\maketitle

\begin{abstract}
As the number of sensors becomes massive in Internet of Things (IoT) networks, the amount of data is humongous.
To process data in real-time while protecting user privacy, federated learning (FL) has been regarded as an enabling technique to push edge intelligence into IoT networks with massive devices.
However, FL latency increases dramatically due to the increase of the number of parameters in deep neural network and the limited computation and communication capabilities of IoT devices.
To address this issue, we propose a semi-federated learning (SemiFL) paradigm in which network pruning and over-the-air computation are efficiently applied.
To be specific, each small base station collects the raw data from its served sensors and trains its local pruned model. 
After that, the global aggregation of local gradients is achieved through over-the-air computation.
We first analyze the performance of the proposed SemiFL by deriving its convergence upper bound.
To reduce latency, a convergence-constrained SemiFL latency minimization problem is formulated.
By decoupling the original problem into several sub-problems, iterative algorithms are designed to solve them efficiently.
Finally, numerical simulations are conducted to verify the effectiveness of our proposed scheme in reducing latency and guaranteeing the identification accuracy.
\end{abstract}


\section{Introduction}
The huge amount of data collected by the increasing number of edge devices in wireless networks has greatly promoted the development of machine learning (ML) technologies \cite{aygun2022hierarchial}.
However, traditional ML, where samples are collected by a server and trained centrally, may not be suitable for Internet of Things (IoT) network.
On the one hand, end devices typically collect samples in private settings, the transmission of raw data poses a great risk of privacy breaches \cite{wu2022node}.
On the other hand, the transmission of massive raw data will increase the burden of network and reduce the communication efficiency \cite{wang2022interference}.
With the increasing prevalence of edge computing and intelligence, the rapid development of distributed ML is considered to have great potential in addressing above issues.
One of the popular distributed ML algorithms is federated learning (FL).
Without revealing their private data to the central server, the participating devices  in FL only use private data for local training.
After that, each device transmits its local parameter over wireless channels to the server for global aggregation \cite{mcmahan2017communication}.
Finally, the edge server updates the global model/gradient and then broadcasts it back to all participating devices.
The above processes are repeated until the global model converges.
Compared with centralized ML, each device needs to communicate with the server more frequently to collaboratively train the FL model \cite{chen2022federated}.
Therefore, it is very important to design efficient wireless resource allocation schemes for FL to support the rapid interaction of model/gradient parameters between end devices and server.

In fact, the joint optimization of learning and communication for wireless FL has been widely studied in \cite{chen2021joint,li2022energy,mo2021energy,ye2020edgefed,zhang2021optimizing,zhang2021energy,lu2021communication}.
Considering the effect of packet transmission error on the performance of global model, the authors of \cite{chen2021joint} proposed an effective algorithm to minimize the training loss by optimizing the transmit power, uplink resource block, and user selection jointly.
In a clustered FL system, Li \textit{et. al} \cite{li2022energy} considered making a trade-off between learning performance and communication cost by optimizing clustering method and resource allocation.
They first proposed an edge association scheme based on deep reinforcement learning (DRL), and then the bandwidth allocation scheme was obtained based on convex optimization.
Furthermore, in an edge server supported FL system, the authors of \cite{mo2021energy} proposed the corresponding resource allocation algorithms under non-orthogonal multiple access (NOMA) and time division multiple access, respectively.
The original problems were transformed firstly, then the authors gave effective solution schemes based on convex optimization.
By integrating edge computing to the wireless FL system, Ye \textit{et. al} \cite{ye2020edgefed} proposed a novel framework, named EdgeFed.
In this framework, split training was first performed on resource-constrained end devices and edge servers, then global model was aggregated in cloud server.
In \cite{zhang2021optimizing}, Zhang \textit{et. al} proposed a deep multi-agent reinforcement learning scheme to minimize the FL loss under FL latency and long-term energy constraints.
The authors of \cite{zhang2021energy} and \cite{lu2021communication} considered digital twin-enabled industrial Internet of Things (IIoT).
Focusing on saving energy, Zhang \textit{et. al} \cite{zhang2021energy} proposed a DRL-based algorithm to optimize the training method selection of IIoT devices and allocation of wireless channels.
While Lu \textit{et. al} \cite{lu2021communication} formulated a FL communication cost minimization problem.
By decomposing the optimization problems in the computation and communication phases, the authors presented the algorithms based on grouping and deep neural networks (DNN), respectively.

\begin{table*}[t!]
	\caption{Literature Comparison}
	\label{comparison}
	\centering
	\begin {tabular}{|c|c|c|c|c|c|}
	\hline
	{}&\textbf{\tabincell{c}{Hierachical\\ network}}&\textbf{\tabincell{c}{Over-the-air\\ computation}}&\textbf{\tabincell{c}{Model\\ compression}}&\textbf{\tabincell{c}{Convergence\\ analysis}}&\textbf{\tabincell{c}{Resource\\ allocation}}\\
	\hline
	\cite{chen2021joint}&{}&{}&{}&\checkmark&\checkmark \\
	\hline
	\cite{li2022energy,zhang2021optimizing,zhang2021energy} &
	\checkmark&{}&{}&{}&\checkmark \\
	\hline
	\cite{mo2021energy}&{}&{}&{}&{}&\checkmark \\
	\hline
	\cite{ye2020edgefed}&\checkmark&{}&{}&{}&{} \\
	\hline
	\cite{lu2021communication}&\checkmark&{}&{}&\checkmark&\checkmark \\
	\hline	
	\cite{zhang2021gradient}&{}&\checkmark&{}&{}&\checkmark \\
	\hline	
	\cite{sun2022dynamic,cao2022transmission,zheng2023semi}&{}&\checkmark&{}&\checkmark &\checkmark \\
	\hline	
	\cite{lin2022relay}&\checkmark&\checkmark&{}&{}&\checkmark \\
	\hline	
	\cite{ni2022semifl}&{}&\checkmark&{}&\checkmark &{} \\
	\hline	
	\cite{yang2021novel,cui2020cluster,ni2023semi}&{}&{}&\checkmark&{}&{} \\
	\hline	
	\cite{park2023regulated,jiang2022model,prakash2022iot,jiang2022fedmp}&{}&{}&\checkmark&\checkmark&{} \\
	\hline	
	\cite{xue2022fedocomp}&{}&\checkmark&\checkmark&\checkmark&{} \\
	\hline	
	\cite{liu2022jointmodel}&{}&{}&\checkmark&\checkmark&\checkmark \\
	\hline	
	\textbf{Our paper}&\checkmark&\checkmark&\checkmark&\checkmark&\checkmark \\
	\hline
\end{tabular}
\vspace{-3mm}
\end{table*}

Since most global aggregation concerns the summation of local weights/gradients \cite{guo2021aggregation}, over-the-air computation supported FL is widely studied in \cite{zhang2021gradient,sun2022dynamic,cao2022transmission,lin2022relay,zheng2023semi,ni2022semifl} to improve the efficiency of FL.
Specifically, Zhang \textit{et. al} \cite{zhang2021gradient} formulated a mean square error (MSE) minimization problem subject to average transmit power constraints.
Subsequently, they proposed a gradient statistics estimation scheme and a transmit power optimization algorithm after obtaining the gradient statistics.
In a multi-user FL system, an energy-constrained training performance optimization problem was formulated in \cite{sun2022dynamic}.
Sun \textit{et. al} first analyzed the convergence rate of the proposed FL scheme and then designed an efficient dynamic device scheduling algorithm.
In the same scenario, the authors of \cite{cao2022transmission} proposed two corresponding power control schemes for power-constrained optimality gap minimization problem and optimality gap-constrained FL latency minimization problem, respectively.
Lin \textit{et. al} \cite{lin2022relay} proposed a relay-assisted over-the-air FL framework and formulated a model aggregation MSE minimization problem.
By decoupling the original problem into serval sub-problems, an alternating algorithm with low time complexity was designed.
Incorporating over-the-air computation, a semi-federated learning framework which integrates centralized learning (CL) and FL has been studied in \cite{zheng2023semi} and \cite{ni2022semifl}.
Specifically, Zheng \textit{et. al} \cite{zheng2023semi} formulated a problem of minimizing the derived convergence upper bound of SemiFL with communication latency and resource budget constraints.
To solve the non-convex problem, an efficient two-stage algorithm was further proposed.
A reconfigurable intelligent surface (RIS) supported SemiFL framework was designed in \cite{ni2022semifl}.
Then the detailed analysis of the effects of learning rate, channel fading and noise on convergence rate of the scheme was given. 
The subsequent numerical results verified that the SemiFL architecture could obtain better learning performance than FL with a latency lower than CL. 
	
In addition to over-the-air computation, the network model compression is also considered as an important approach to alleviate the contradiction between the huge data volume of model and the limited resources in the wireless network.
To this end, it has received extensive attention from researchers in \cite{yang2021novel,cui2020cluster,park2023regulated,xue2022fedocomp}.
In \cite{yang2021novel}, Yang \textit{et. al} proposed an adaptive gradient compression scheme to improve communication efficiency.
In above scheme, the training state perception algorithm is first executed, before the compression ratio is calculated based on perception data.
Then, each client selects partial gradients based on their importance to upload according to the obtained compression ratio. 
Finding that only a small fraction of the gradients are far from zero in each communication round, Cui \textit{et. al} \cite{cui2020cluster} designed a clustering-based gradient compression algorithm.
Specifically, the K-means clustering algorithm was applied to cluster all gradients.
Then quantization schemes for the gradients in different clusters were given.
Considering the correlation of model updates, Park \textit{et. al} \cite{park2023regulated} proposed a model update compression scheme based on subspace projection.
For further improving the accuracy of global model, a criteria which determines whether each local model should be compressed was given.
By saving communication resources with over-the-air computation, Xue \textit{et. al} \cite{xue2022fedocomp} proposed an online gradient compression scheme, named FedOComp, which utilizes the correlation of gradient structure.

However, with the widespread application of DNN, training millions of parameters will consume plenty of storage space and computational resource \cite{knight2020performance}.
In this case, efficient training of the high-dimension models on end devices with limited computational resources becomes extremely challenging.
To address above issue, the adaptive network pruning which reduces model size by removing part of the least important weights in DNN is considered in \cite{jiang2022model,prakash2022iot,jiang2022fedmp,liu2022jointmodel,ni2023semi}.
In \cite{jiang2022model}, a network pruning supported FL scheme (PruneFL), which contains both initial and further pruning process was designed.
In view of the limited resources of end devices in IoT scenarios, Prakash \textit{et al.} \cite{prakash2022iot} proposed a novel FL framework that utilizes both quantization and network pruning.
Subsequently, its convergence was demonstrated through theoretical analysis and simulation.
The authors of \cite{jiang2022fedmp} designed a pruning rates optimization algorithm based on multi armed bandit for a model pruning supported FL system.
In \cite{liu2022jointmodel}, Liu \textit{et. al} first gave the expression of the FL convergence rate.
Then a joint device selection, pruning rate and time fraction allocation algorithm was proposed to speed up the convergence of global model.
A SemiFL framework considering client selection and network pruning was proposed in \cite{ni2023semi}, in which IoT devices with different hardware abilities could participate in the learning process equally.

The detailed comparison between this paper and other literatures is presented in Table \ref{comparison}.
It can be found that although literatures \cite{chen2021joint,li2022energy,mo2021energy,ye2020edgefed,zhang2021optimizing,zhang2021energy,lu2021communication} proposed some resource allocation schemes for efficient FL, the extremely high communication and computational load caused by the dramatic growth of model data amount greatly hindered their deployment in low-latency scenarios.
To address above issue, literatures \cite{zhang2021gradient,sun2022dynamic,cao2022transmission,lin2022relay,zheng2023semi,ni2022semifl} studied the over-the-air computation supported FL to speed up the model/gradient uploading and global aggregation process.
At the same time, literatures \cite{yang2021novel,cui2020cluster,park2023regulated,xue2022fedocomp} applied model compression technology to reduce the data amount of gradient/model that needs to be uploaded by devices.
Literatures \cite{zhang2021gradient,sun2022dynamic,cao2022transmission,lin2022relay,zheng2023semi,ni2022semifl,yang2021novel,cui2020cluster,park2023regulated,xue2022fedocomp} focused on improving communication efficiency during the model uploading process after each local model was obtained.
However, still using the full model for training made all above schemes with no help to accelerate the local model training.
This greatly affected the effectiveness of the above algorithms, especially when the computing capabilities of devices in wireless network were limited.
With aim to accelerating the local training process of each device, the resource optimization algorithms for network pruning supported FL were proposed in \cite{jiang2022model,prakash2022iot,jiang2022fedmp,liu2022jointmodel,ni2023semi}.
However, compared with over-the-air computation, the orthogonal multiple access applied still greatly limited the model/gradient uploading rate. 
Besides, the decoding and aggregation of the received signal also increased communication overhead as well as the FL latency.	

\subsection{Motivations and Challenges}
Based on Table \ref{comparison} and above analysis, we can find that the framework that can comprehensively utilize the advantages of network pruning and over-the-air computation in accelerating model training and parameter uploading is of great potential, especially for the IoT networks.
However, its design still faces some challenges as follows.
\begin{itemize}
\item Firstly, the low-cost sensors are the main devices for sensing in future IoT, but it is often difficult for them to complete the FL local training because of their energy budget and limited computing capacities.
Therefore, there is an urgent need to design a FL framework that can effectively help IoT devices to process data samples and train local models.
\item Secondly, the performance of FL will be affected by the local pruning rates in network pruning and the deviation between the estimated and desired signal during over-the-air computation. 
However, the way to properly characterize the multiple effects on FL convergence rate is extremely challenging.
\item Thirdly, the coupling of wireless resources and local pruning rates on FL latency and learning performance greatly increases the difficulty for designing efficient joint resource allocation algorithm.
\end{itemize}
\subsection{Contributions and Organization}
To overcome the above challenges, a SemiFL framework that effectively integrates network pruning and over-the-air computation is proposed.
The main contributions of this paper are summarized as follows:
\begin{itemize}
	\item[1)] We propose a SemiFL framework for resource-constrained IoT networks.
	Within one round, each small base station (SBS) first collects samples uploaded by its served sensors through non-orthogonal multiple access (NOMA).
	Subsequently, each SBS prunes the latest global model and starts the local training of pruned model.
	Finally, after aggregating all local gradients with over-the-air computation at the macro base station (MBS), the global model is updated and broadcasted. 
	\item[2)] To facilitate analysis, we derive the convergence upper bound of proposed SemiFL under the assumptions of non-convex loss function and non-IID data. 
	The derivation result reveals the effects of data heterogeneity, power control mechanism, wireless channel noise and network pruning rates on the convergence rate of SemiFL.
	\item[3)] 
	Based on the derived convergence upper bound, we first formulate a long-term latency minimization problem.
	By transforming the constraint of long-term convergence upper bound, the original problem is simplified to a single-round problem.
	Finally, it is further decoupled into several sub-problems, before the corresponding efficient algorithms are designed to solve them respectively.

	\item[4)] We verify the performance of our proposed algorithm by numerical simulations.
	The numerical results demonstrate the effectiveness of our algorithm in reducing FL latency and ensuring identification accuracy.
	
\end{itemize}

The remainder of this paper is organized as follows.
In Section \Rmnum{2}, the system model and the communication model of proposed SemiFL are given.
The convergence upper bound of proposed SemiFL is derived and the original optimization problem is given in Section \Rmnum{3}.
To solve this problem, an efficient iterative algorithm is proposed in Section \Rmnum{4}. 
Finally, we present the numerical results in Section \Rmnum{5}, which is followed by conclusion and future work in Section \Rmnum{6}.

\section{System Model}
	\label{system}
\begin{figure*} [t!]
	\centering
	\includegraphics[width=5.4 in]{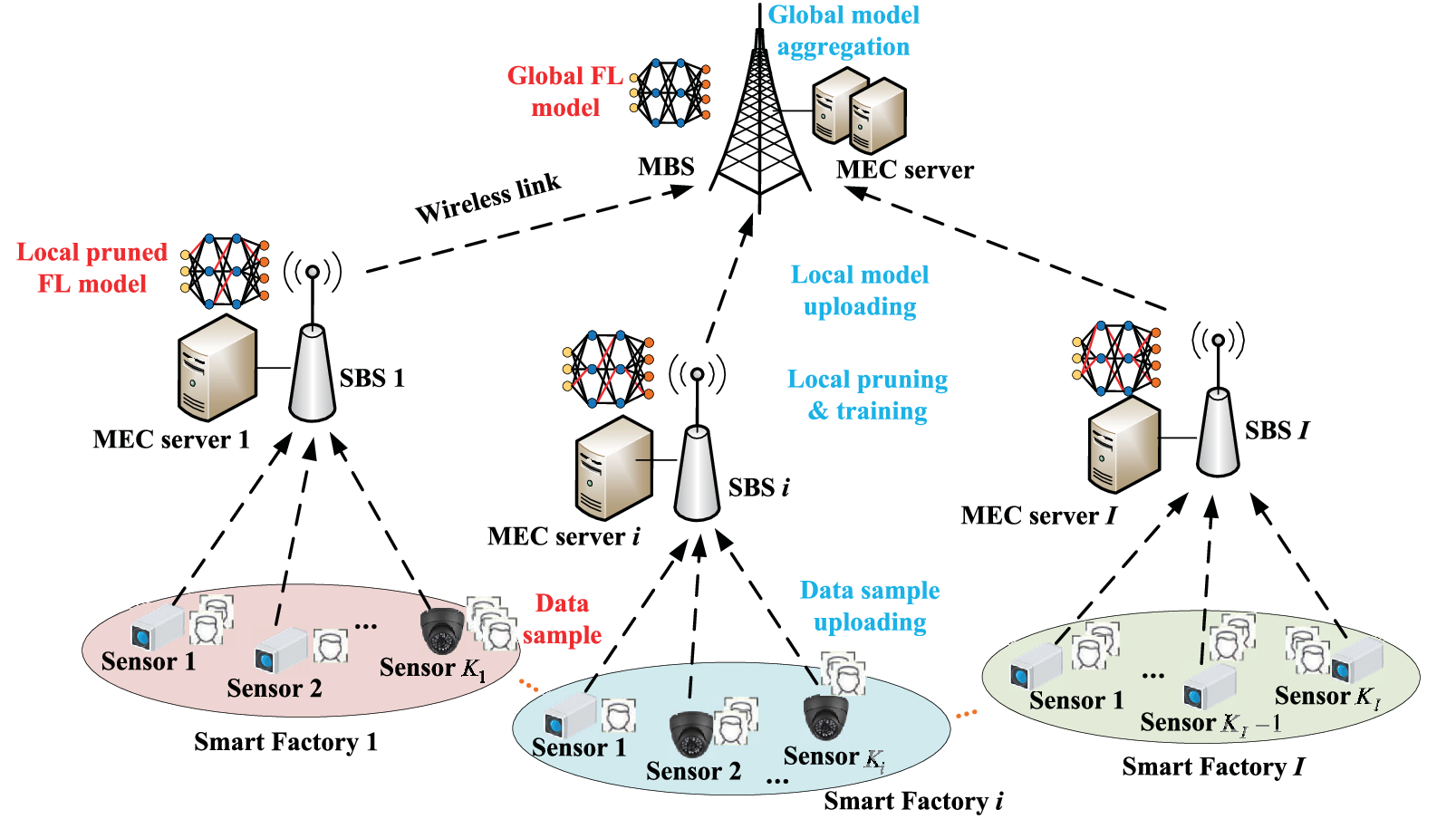}
	\caption{An illustration of the proposed SemiFL networks}
	\label{system_model}
\end{figure*}
As shown in Fig. \ref{system_model}, we consider an IoT network consisting of one MBS and $I$ SBSs, denoted by $\mathcal{I} = \{ 1,2, \ldots, I\}$.
The set of sensors served by SBS $i$ is indexed by $\mathcal{S}_{i}$ and $N_i=|\mathcal{S}_{i}|$ is the number of sensors served by SBS $i$. 
In each round, each SBS selects a portion of sensors within its service range to upload their samples.
In order to improve the spectral efficiency in the uploading process, the NOMA is considered for sensors' uplink transmission.
To avoid the inter-cell interference, each SBS monopolizes a frequency band with bandwidth $B$. 

Each communication round of our proposed SemiFL is composed of three parts: 1) sample collection, 2) distributed training, 3) gradient uploading and global aggregation.

\subsubsection{Sample collection}
In SBS $i$, let $\hat{g}_{k,i}, (1\le k\le N_i)$ denote the complex uplink channel from sensor $k$ to SBS $i$, and $g_{k,i}=|\hat{g}_{k,i}|$ is its magnitude.
Without loss of generality, we assume the channel gains of the sensors served by SBS $i$ are sorted in a non-increasing order, i.e. $g_{1,i}^2\ge g_{2,i}^2\ge\ldots\ge g_{N_i,i}^2$.
Let $c_{k,i}\in\{0,1\}$ denote the sensor selection indicator.
If sensor $k$ in SBS $i$ is selected to upload its samples, $c_{k,i}=1$, otherwise, $c_{k,i}=0$.
Let $x_{k,i}$ denote its transmit symbol which is normalized with unit variance.
Then with proper phase control of all sensors, the signal received by SBS $i$ is represented as:
\begin{equation}
	\setlength\abovedisplayskip{2pt}
	\setlength\belowdisplayskip{2pt}
	y_i=\sum\nolimits_{k\in \mathcal{S}_i}c_{k,i}g_{k,i}{p}_{k,i}x_{k,i}+z,
\end{equation}
where $p_{k,i}\ge 0$ denote power control factor of sensor $k$ in cell $i$ and $z$ is the additional white Gaussian noise with power $n_0$.

Based on the successive interference cancellation (SIC) technology, where the received signals are iteratively decoded according to the order of channel gains, the interference remains when decoding the signal of sensor $k$ in cell $i$ can be represented as
\begin{equation}\label{interference_remain}
	\setlength\abovedisplayskip{2pt}
	\setlength\belowdisplayskip{2pt}
	{I_{k,i}} =\sum\nolimits_{j=k+1}^{N_i}c_{j,i}g_{j,i}^2p_{j,i}^2.
\end{equation}
With (\ref{interference_remain}), the achievable uplink transmission rate of sensor $k$ can be expressed as
\begin{equation}\label{uplink_rate_sensor}
	\setlength\abovedisplayskip{2pt}
	\setlength\belowdisplayskip{2pt}
	{R_{k,i}} =B{\rm log}_2\left(1+\frac{g_{k,i}^2p_{k,i}^2}{I_{k,i}+n_0}\right).
\end{equation}

Let $N_{k,i}$ denote the number of local data samples transmitted by sensor $k$ in cell $i$ and $D_{\rm s}$ denote the data size of each sample. Then the transmission latency $t_{k,i}$ is characterized by
\begin{equation}\label{uplink_latency_sensor}
	\setlength\abovedisplayskip{2pt}
	\setlength\belowdisplayskip{2pt}
	{t_{k,i}} =\left\{
	\begin{array}{lll}
		\!0,&&\text{if}\ c_{k,i}=0,\\
		\!{N_{k,i}D_{s}}/{R_{k,i}},&&\text{if}\ c_{k,i}=1.
	\end{array}\right.
\end{equation}

\subsubsection{Distributed training}
When all data samples are collected, SBS $i$ performs the local pruning of whole model and then starts its distributed training before uploading it for aggregation.
Let $\tilde{W}_i^t=\mathbf{f}_{i}^{\rm P}\left(W^t\right)$ denote the local pruned model at SBS $i$ at the $t$-step, where function $\mathbf{f}_i^{\rm p}(\cdot)$ represents the pruning operation and $W^t$ is its latest received global model.
To facilitate analysis, we denote the pruning rate as $\rho_i=D_i^{\rm p}/D_{\rm M}$, where $D_i^{\rm p}$ and $D_{\rm M}$ are the data size pruned by SBS $i$ and the data size of global model $W^t$, respectively.
Thus the local training latency at SBS $i$ can be represented by
\begin{equation}\label{train_latency_sbs}
	\setlength\abovedisplayskip{2pt}
	\setlength\belowdisplayskip{2pt}
	{t_{i}^{\rm l}} ={\left(1-\rho_i\right)d^{\rm c}K_i}/{f_i},
\end{equation} 
where $K_i=\sum\nolimits_{k=1}^{N_i}c_{k,i}N_{k,i}$ is the total number of samples collected by SBS $i$, $d^{\rm c}$ represents the central process unit (CPU) cycles required to compute one sample and $f_i$ is the allocated CPU cycles per second.
To this end, the total latency from data collection to the completion of local training at SBS $i$ can be expressed as
\begin{equation}\label{total_latency_dc_cc}
	\setlength\abovedisplayskip{2pt}
	\setlength\belowdisplayskip{2pt}
	{t_{i}^{\rm c}} =\max\nolimits_{k\in \mathcal{S}_i}\{t_{k,i}\}+t_i^{\rm l}.
\end{equation} 

\subsubsection{Gradient uploading and global aggregation}
Over-the-air computation is applied to support the fast global gradient aggregation.
Let $x_i$ denote the transmit symbol of the SBS $i$, which is assumed to be normalized with unit variance.
Furthermore, we assume that $\mathbb{E}(x_ix_j)=0, \forall i\neq j$, which indicates the information symbols of different SBSs are statistically independent.
Therefore, we can express the signal received at the MBS (after phase compensation) as
\begin{equation}\label{received_signal}
	\setlength\abovedisplayskip{2pt}
	\setlength\belowdisplayskip{2pt}
	y=\sum\nolimits_{i=1}^Ig_i{\overline{p}_i}x_i+z_0,
\end{equation}
where $g_i$ is the magnitude of uplink channel from SBS $i$ to the MBS, $\overline{p}_i\ge 0$ is the power control factor of SBS $i$ and $z_0\sim\mathcal{N}(0,\sigma^2)$ is the noise.
Given the maximum power budget $p^{\rm max}$, the power control factor of SBS $i$ is constrained by
\begin{equation}\label{power_constraint}
	\setlength\abovedisplayskip{2pt}
	\setlength\belowdisplayskip{2pt}
	\mathbb{E}\left(({\overline{p}_i}x_i)^2\right)={\overline{p}_i}^2\le p^{\rm max}.
\end{equation}

Upon receiving the signal, its estimation obtained at the MBS can be given by
\begin{equation}\label{estimated_signal}
	\setlength\abovedisplayskip{2pt}
	\setlength\belowdisplayskip{2pt}
	\hat{x}=ay=\sum\nolimits_{i=1}^Iag_i{\overline{p}_i}x_i+az_0,
\end{equation}
where $a$ denotes post-transmission factor.
Considering the desired signal $x=\frac{1}{K}\sum_{i=1}^IK_ix_i,\ (K=\sum_{i=1}^IK_i)$, we can express the aggregation distortion as
\begin{equation}\label{aggregation_distortion}
	\setlength\abovedisplayskip{2pt}
	\setlength\belowdisplayskip{2pt}
	{\rm MSE}\left(\hat{x},x\right)=\frac{1}{K^2}\sum\nolimits_{i=1}^I\left(Kag_i{\overline{p}_i}-K_i\right)^2+a^2\sigma^2.
\end{equation}
Considering that the above MSE can be regarded as the interference-plus-noise power of over-the-air computation, further according to \cite{hu2021energy,ni2022integrating}, the computation rate of model aggregation can be represented as
\begin{equation}\label{aggregation_rate}
	\setlength\abovedisplayskip{2pt}
	\setlength\belowdisplayskip{2pt}
	R^{\rm a}=B_{\rm M}{\rm log}_2\left(1+\frac{\mathbb{E}\left(\hat{x}^2\right)-{\rm MSE}\left(\hat{x},x\right)}{{\rm MSE}\left(\hat{x},x\right)}\right),
\end{equation}
where $B_{\rm M}$ is the bandwidth of MBS.
Then the latency of global gradient aggregation can be given by $t^{\rm a}=D_{\rm M}/R^{\rm a}$.

After recovering the local stochastic gradients {$G_{i}^{t}$} from wireless signals {$x_i$} at the MBS, the estimated global aggregated gradient can be shown as
\begin{equation}\label{global_gradient}
	\setlength\abovedisplayskip{2pt}
	\setlength\belowdisplayskip{2pt}
	\hat{G}^t=\sum\nolimits_{i=1}^Iag_i^t{\overline{p}_i^t}G_i^{t}+a\sigma_{\rm n}^t\mathbf{z}_0,
\end{equation} 
where $\sigma_{\rm n}^t$, same as $\pi(t)$ in \cite{zou2023knowledge}, denotes the element-wise standard deviation coefficient applied in previous normalization process, the $\mathbf{z}_0\in \mathbb{R}^Q$ and $Q$ is the dimension of gradient.
With $\hat{G}^t$, the global model update rule can be written as 
\begin{equation}\label{global_model_update_rule}
	\setlength\abovedisplayskip{2pt}
	\setlength\belowdisplayskip{2pt}
	W^{t+1}=W^t-\eta\hat{G}^t,
\end{equation} 
where $\eta$ denotes the learning rate.

Because a large transmission power and full bandwidth can be used at MBS for downlink transmission, the latency of global model broadcasting is neglected in this paper. 
To sum up, the total latency for completing one communication round of proposed SemiFL is given by
\begin{equation}\label{total_latency}
	\setlength\abovedisplayskip{2pt}
	\setlength\belowdisplayskip{2pt}
	T^{\rm one}=\max\nolimits_{i\in\mathcal{I}}\left\{t_i^{\rm c}+t^{\rm a}\right\}.
\end{equation} 

The above system model can be effectively applied in the Internet of Vehicles (IoV) \cite{zhou2021two}, the Industrial Internet of Things (IIoT) \cite{ji2023joint}, and the Internet of Agricultural Things (IoAT) \cite{yu2022energy}.
Specifically, for the road sign recognition and collision prediction in IoV, the wireless sensors in each vehicle first transmit their collected sample to the vehicle's central server.
Each server trains its local model and then uploads local parameters to the roadside unit (RSU) for global aggregation.
As for the product quality monitoring and accident detection in the IIoT, the samples perceived by low-cost wireless sensors can be collected and utilized for local training by edge servers. 
Subsequently, all local weights/gradients are aggregated at the cloud server.
Similarly, for applications such as pest/plant disease diagnosis in IoAT, unmanned aerial vehicle (UAV) swarm can be regularly used to collect samples such as crop images, which are then transmitted to farm edge nodes to establish a sample database and undergo local model training.
Local models from different farms can be aggregated at MBS to improve the performance of the global FL model.

\section{Convergence Analysis and Problem Formulation}
\subsection{Convergence Analysis}
In actual network, the data samples collected by different SBSs are usually non-independent and identically distributed (Non-IID).
To this end, we analyze the convergence rate of proposed SemiFL under Non-IID datasets in this part.
To facilitate analysis, the following assumptions are given at first:

\noindent\textbf{Assumption 1.} Loss function $F\left(\cdot\right)$ is $\beta$-smooth \cite{liu2022jointmodel}:
\begin{equation}
	\setlength\abovedisplayskip{2pt}
	\setlength\belowdisplayskip{2pt}
	F\left(V\right)\!\leq\!{F\left(W\right)}\!+\!\left<V-W,\nabla{F\left(W\right)}\right>\!+\!
	\frac{\beta}{2}{\left\|V-W\right\|^2}.
\end{equation} 

\noindent\textbf{Assumption 2.} The model weight and local stochastic gradient $G_{i}^{t}\ (\forall i,t)$ can always be bounded by non-negative constants $D$ and $M$ respectively \cite{liu2022jointmodel}, thus we have 
\begin{equation}
	\setlength\abovedisplayskip{2pt}
	\setlength\belowdisplayskip{2pt}
	\mathbb{E}\left[\|W^t\|^2\right]\le D^2,\ \forall t,
\end{equation} 
\begin{equation}
	\setlength\abovedisplayskip{2pt}
	\setlength\belowdisplayskip{2pt}
	\mathbb{E}\left[\|G_{i}^{t}\|^2\right]\le M^2,\ \forall i,t.
\end{equation}

\noindent\textbf{Assumption 3.} The stochastic gradient calculated with $\tilde{W}_i^t$ on data sample $j$ is unbiased and variance-bounded, that is
\begin{equation}
	\setlength\abovedisplayskip{2pt}
	\setlength\belowdisplayskip{2pt}
	\mathbb{E}\left[G_{i,j}^{t}\right]= \nabla{F_i\left(\tilde{W}_i^t\right)},
\end{equation}
\begin{equation}
	\setlength\abovedisplayskip{2pt}
	\setlength\belowdisplayskip{2pt}
	\mathbb{E}\left[\|\nabla{F_i\left(\tilde{W}_i^t\right)}-G_{i,j}^{t}\|^2\right]\le \phi^2,\ \forall i,j,t,
\end{equation}
where above expectation is with respect to stochastic data sampling and $\nabla{F_i\left(\tilde{W}_i^t\right)}$ is the gradient of loss function at SBS $i$ \cite{cao2022transmission,wan2021convergence}.

\noindent\textbf{Assumption 4.} The gradient divergence between local and global gradient are bounded by $U\ge 0$:
\begin{equation}
	\setlength\abovedisplayskip{2pt}
	\setlength\belowdisplayskip{2pt}
	\|\nabla{F\left(W\right)}-\nabla{F_i\left(W\right)}\|\le U,\ \forall i\in \mathcal{I},
\end{equation}
where $U$ reflects the heterogeneity between SBSs' local datasets \cite{cao2022transmission}.

Based on above assumptions, the convergence upper bound of proposed SemiFL which is defined as the average $l_2$-norm of its global gradient can be given in the following theorem.

\begin{theorem}\label{theorem_convergence_rate}
	With $\sum_{i=1}^I(K_i^t-K^tag_i^t{\overline{p}_i^t})^2\leq b(K^t)^2\ (\forall t)$, the average $l_2$-norm of global gradient which reflects the expected convergent rate of SemiFL is given by
	\begin{align}\label{convergence_rate}
			&\frac{1}{S+1}\sum_{t=0}^S\mathbb{E}\left[\|\nabla{F\left({W}^t\right)}\|^2\right]
			\le\frac{{F\left(W^0\right)}-F\left(W^*\right)}{\left(S+1\right)/(2\beta)}+d\\ \notag&\ \ \ \!+\!\mathbb{E}_t\!\left[\!\frac{6\beta^2 ID^2}{(K^t)^2}\sum_{i=1}^I(K_i^t)^2\rho_i^t\!+\!\frac{6IU^2}{ (K^t)^2}\sum_{i=1}^I(K_i^t)^2+\frac{6I\phi^2}{K^t}\right],
	\end{align}
    where $d=2(Q{a^{\prime}}^2\sigma^2+bIM^2)$ and $\mathbb{E}_t\left[\cdot\right]$ computes the average value during $t=0$ to $t=S$.
\end{theorem}

\begin{IEEEproof}
	Please see Appendix A.
\end{IEEEproof}

From Theorem \ref{theorem_convergence_rate}, we can find that many factors, such as total number of samples, data heterogeneity, power factor control mechanism, noise power and network pruning rates affect the convergence upper bound of SemiFL.
Specifically, higher noise power, higher local pruning rates and more prominent data heterogeneity increase the upper bound of the average $l_2$-norm of global gradients, that is slowing down its convergence rate.
This is consistent with the actual situation, because higher noise power brings higher aggregation distortion.
While higher local pruning rates and more prominent data heterogeneity increase the deviation between the local model and the global model.
Both aggregation distortion and the deviation between the local model and the global model reduces the convergence rate of the FL global model. 
Besides, with higher the requirement for the MSE of the signal, that is, the smaller $b$, the lower the convergence upper bound is.
Finally, when the proportion of each SBS' local samples to the total sample size remains unchanged, the larger the total sample number is, the faster global model converges.

\begin{figure*} [t!]
	\centering
	\includegraphics[width=6.4 in]{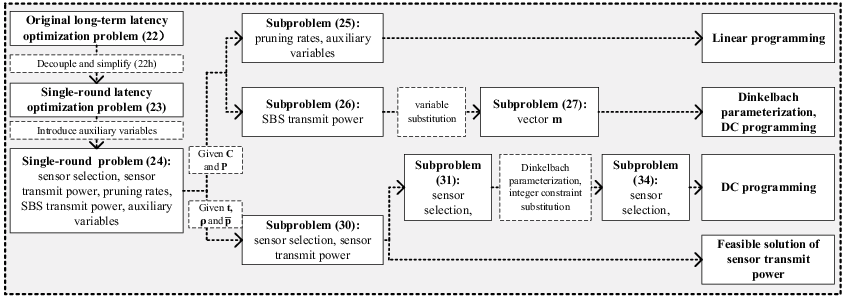}
	\vspace{-2mm}
	\caption{An overview of problem decomposition and corresponding algorithms}
	\label{overview}
\end{figure*}

\subsection{Problem Formulation}
In addition to the convergence upper bound of SemiFL, its latency is also an important indicator in many scenarios, such as autonomous driving, smart manufacturing, and smart medical care.
To support the practical implement of SemiFL, both its total latency in (\ref{total_latency}) and its convergence behavior in (\ref{convergence_rate}) should be optimized.
Further considering the effects of the power factor control mechanism, sensor selection and network pruning rates on above two metrics, they should be carefully designed.
Based on the description of single-round SemiFL above, we have further added an identification $t$ to indicate which round of communication it is.
Specifically, let $T_t^{\rm{one}}$ denote the total latency of the $t$-th communication round, other variables such as $\rho_i^t$, $c_{k,i}^t$, $K_i^t$, etc. are similar.
Overall, with aim to reduce the total latency of proposed SemiFL while guaranteeing its convergence rate, our optimization problem is:
\begin{subequations}\label{problem_long_term_total_latency_minimization}
	\begin{eqnarray}
		&\min \limits_{ \boldsymbol{\rho}_{\rm{l}}, \mathbf{c}_{\rm{l}}, \mathbf{p}_{\rm{l}},\mathbf{\overline{p}}_{\rm{l}}} & \sum\nolimits_{t=0}^{S} T_t^{\rm one} \\
		\label{total_latency_minimization_prune_constraint_long}
		&{\rm s.t.}&\rho_i^{\rm min} \le \rho_i^t \le \rho_i^{\rm max},\ \forall i,t\\
		\label{total_latency_minimization_c_constraint_long}
		&{}&c_{k,i}^t\in\{0,1\},\ \forall k,i,t, \\
		\label{total_latency_minimization_k_constraint_long}
		&{}&K_i^t\ge K_i^{\rm min},\ \forall i,t, \\
		\label{total_latency_minimization_sensor_power_constraint_long}
		&{}&0 \le p_{k,i}^t \le \sqrt{p^{\rm sensor}},\ \forall k,i,t, \\
		\label{total_latency_minimization_sbs_power_constraint_long}
		&{}&0 \le {\overline{p}_i^t} \le \sqrt{p^{\rm max}}, \forall i,t, \\
		\label{total_latency_minimization_mse_constraint_long}
		&{}& \sum\nolimits_{i=1}^I(K_i^t-K^tag_i^t{\overline{p}_i^t})^2\le b(K^t)^2,\forall t, \\
		\label{long_term_convergence_constraint}
		&{}&\!m\!\mathbb{E}_t\!\left[\!\frac{\!\sum\nolimits_{i}(K_i^t)^2\rho_i^t\!+\!\!\sum\nolimits_{i}\!(K_i^t)^2}{(K^t)^2}\!+\!\!\frac{1}{K^t}\!\right]\!\le\!\xi,
	\end{eqnarray}
\end{subequations}
where $m=\max\{\beta^2D^2,U^2,\phi^2\}$ and $\xi\textgreater0$ represents the pre-set convergence upper bound.
The variable $m$ unifies the coefficients of each term in (\ref{convergence_rate}), effectively avoiding the cumbersome exploration process of various parameter values in the assumptions.
Constraint (\ref{total_latency_minimization_prune_constraint_long}) limits the local pruning rates, where $\rho_i^{\rm min}$ and $\rho_i^{\max}$ are the minimal and maximal local pruning rates at SBS $i$. 
They are related to storage limitations and acceptable performance degradation, respectively.
Then constraint (\ref{total_latency_minimization_c_constraint_long}) is the value range constraint to $\mathbf{C}$ and (\ref{total_latency_minimization_k_constraint_long}) denotes the number of collected samples at SBS $i$ should be greater than the minimum sample number $K_i^{\rm min}$.
In (\ref{total_latency_minimization_sensor_power_constraint_long}) and (\ref{total_latency_minimization_sbs_power_constraint_long}), $p^{\rm sensor}$ and $p^{\rm max}$ are respectively the maximum transmit power of each sensor and SBS.
Finally, the constraints of each-round aggregation distortion and long-term convergence upper bound are given in (\ref{total_latency_minimization_mse_constraint_long}) and (\ref{long_term_convergence_constraint}).
Problem (\ref{problem_long_term_total_latency_minimization}) is a long-term optimization problem which is really hard to solve, because addressing it requires to predict the channel state information for a long time in the future.
To overcome above issue, we first replace constraint (\ref{long_term_convergence_constraint}) with a stricter constraint $m\max_t\!\left\{\!\frac{\!\sum\nolimits_{i}(K_i^t)^2\rho_i\!+\!\sum\nolimits_{i}\!(K_i^t)^2}{(K^t)^2}\!+\!\frac{1}{K^t}\right\}\!\le\!\xi$ based on the fact that the mean value of a group of data is less than the maximum value among them.

Then for any round, the constraint we need to consider is $m(\frac{\!\sum\nolimits_{i}(K_i^t)^2\rho_i\!+\!\!\sum\nolimits_{i}\!(K_i^t)^2}{(K^t)^2}\!+\!\frac{1}{K^t})\le\!\xi,(\forall t)$.
Further with $K_i^t\le K^t, (\forall i)$, we define the single-round convergence upper bound of proposed SemiFL as $m(\frac{1}{K}\sum\nolimits_{i=1}^IK_i(\rho_i+1)+\frac{1}{K})$.
Then we can formulate the single-round optimization problem as:
\begin{subequations}\label{problem_total_latency_minimization}
	\begin{eqnarray}
		\label{total_latency_minimization_objective}
		&\min \limits_{ \boldsymbol{\rho}, \mathbf{c}, \mathbf{p},\mathbf{\overline{p}}} & T^{\rm one} \\
		\label{total_latency_minimization_prune_constraint}
		&{\rm s.t.}&\rho_i^{\rm min} \le \rho_i \le \rho_i^{\rm max},\ \forall i,\\
		\label{total_latency_minimization_c_constraint}
		&{}&c_{k,i}\in\{0,1\},\ \forall k,i, \\
		\label{total_latency_minimization_k_constraint}
		&{}&K_i\ge K_i^{\rm min},\ \forall i, \\
		\label{total_latency_minimization_sensor_power_constraint}
		&{}&0 \le p_{k,i} \le \sqrt{p^{\rm sensor}},\ \forall k,i, \\
		\label{total_latency_minimization_sbs_power_constraint}
		&{}&0 \le {\overline{p}_i} \le \sqrt{p^{\rm max}}, \forall i, \\
		\label{total_latency_minimization_mse_constraint}
		&{}& \sum\nolimits_{i=1}^I(K_i-Kag_i{\overline{p}_i})^2\le bK^2, \\
		\label{total_latency_minimization_convergence_constraint}
		&{}&\frac{m}{K}\sum\nolimits_{i=1}^IK_i(\rho_i+1)+\frac{m}{K}\le \xi,
	\end{eqnarray}
\end{subequations}
where $\mathbf{c}=[c_{1,1},\ldots,c_{N_1,1},c_{1,2},\ldots,c_{N_2,2},c_{1,3},\ldots,c_{N_I,I}]^{\rm{T}}$, $\mathbf{p}=[p_{1,1},\ldots,p_{N_1,1},\ldots,p_{N_I,I}]^{\rm{T}}$, $\boldsymbol{\rho}=[\rho_1,\rho_2,\ldots, \rho_I]^{\rm{T}}$ and $\mathbf{\overline{p}}=[\overline{p}_1,\ldots,\overline{p}_I]^{\rm{T}}$. 
Constraint (\ref{total_latency_minimization_convergence_constraint}) is the single round convergence upper bound transformed from the long-term constraint (\ref{long_term_convergence_constraint}).

By replacing (\ref{long_term_convergence_constraint}) with more stringent constraint (\ref{total_latency_minimization_convergence_constraint}), the original long-term latency minimization problem can be simplified into the single-round problem.
However, due to the non-convexity of (\ref{total_latency_minimization_objective}), (\ref{total_latency_minimization_mse_constraint}), (\ref{total_latency_minimization_convergence_constraint}) and the existence of integer optimization variable $\mathbf{c}$, problem (\ref{problem_total_latency_minimization}) is a mixed integer nonlinear programming (MINLP), which is still difficult to solve directly.
To this end, we decouple it to several sub-problems and design corresponding algorithms to solve them in the following section.
An overview of the detailed problem decomposition process and the corresponding algorithms are shown in Fig. \ref{overview}.
Note that the proposed algorithm can effectively solve problem (\ref{problem_total_latency_minimization}), thus give a feasible and sub-optimal solution for original problem (\ref{problem_long_term_total_latency_minimization}) with low complexity.

\section{Proposed Solution}
In this section, we first introduce some auxiliary variables to the original problem and further decouple it into several sub-problems.
Specifically, given $\mathbf{c}$ and $\mathbf{p}$, the original problem can be decomposed into two independent sub-problems which can be solved with convex optimization tools and Dinkelbach parameterization technique-based difference of convex functions (DC) programming, respectively. 
Next, for the joint optimization problem of $\mathbf{c}$ and $\mathbf{p}$, we optimize $\mathbf{c}$ at first. 
By relaxing the integer constraints of $\mathbf{c}$ to continuous ones, this problem can be solve with the Dinkelbach parameterization technique-based DC programming as well.
Finally, we manage to find one feasible power control factor solution $\mathbf{p}$ which satisfies its corresponding constraints.

With the introduced auxiliary variables $T_i\ (i=1,\ldots,I)$ and $T$, the original problem (\ref{problem_total_latency_minimization}) can be reformulated as
\begin{subequations}\label{reformulated_problem_total_latency_minimization}
	\begin{eqnarray}
		\label{re_total_latency_minimization_objective}
		&\min \limits_{ \mathbf{t},\boldsymbol{\rho}, \mathbf{c}, \mathbf{p},\mathbf{\overline{p}}} & \!T\!+\!\frac{D_{\rm M}/B_{\rm M}}{{\rm log}_2\!\left(\!\frac{\sum_i(ag_i\overline{p}_i)^2+a^2\sigma^2}{\left(\sum_i\!(Kag_i\overline{p}_i-K_i)^2\right)\!/\!K^2+a^2\sigma^2}\!\right)} \\
		\label{re_total_latency_minimization_ct_constraint}
		&{\rm s.t.}&t_{k,i}\le T_i,\ \forall k,i,\\
		\label{re_total_latency_minimization_cct_constraint}
		&{}&T_i\!+\!\frac{\left(1-\rho_i\right)d^{\rm c}\!\sum_{k=1}^{N_i}\!c_{k,i}\!N_{k,i}}{f_i}\!\le\!T,\ \forall i,\\
		&{}&(\ref{total_latency_minimization_prune_constraint})-(\ref{total_latency_minimization_convergence_constraint}),
	\end{eqnarray}
\end{subequations}
where $\mathbf{t}=[T_1,\ldots,T_I,T]^{\rm{T}}$.
Then, we further decouple the problem (\ref{reformulated_problem_total_latency_minimization}) into several sub-problems and give effective solution for each sub-problem.

Firstly, when $\mathbf{c}$ and $\mathbf{p}$ are given, problem (\ref{reformulated_problem_total_latency_minimization}) can be decoupled into two sub-problems as  
\begin{subequations}\label{problem_T_minimization}
	\begin{eqnarray}
		(\mathcal{P}1): 
		\label{T_minimization_objective}
		&\min \limits_{ \mathbf{t},\boldsymbol{\rho} } & T \\
		&{\rm s.t.}&(\ref{total_latency_minimization_prune_constraint}),(\ref{total_latency_minimization_convergence_constraint}),(\ref{re_total_latency_minimization_ct_constraint}),(\ref{re_total_latency_minimization_cct_constraint}),
	\end{eqnarray}
\end{subequations}
and
\begin{subequations}\label{problem_uploading_latency_minimization}
	\begin{eqnarray}
		(\mathcal{P}2): 
		\label{ut_minimization_objective}
		&\min \limits_{ \mathbf{\overline{p}}} & \frac{D_{\rm M}/B_{\rm M}}{{\rm log}_2\!\left(\!\frac{\sum_i(ag_i\overline{p}_i)^2+a^2\sigma^2}{\left(\sum_i(Kag_i\overline{p}_i-K_i)^2\right)/K^2+a^2\sigma^2}\!\right)} \\
		&{\rm s.t.}&(\ref{total_latency_minimization_sbs_power_constraint})\ {\rm and}\ (\ref{total_latency_minimization_mse_constraint}).
	\end{eqnarray}
\end{subequations}
It can be found that problem (\ref{problem_T_minimization}) is a linear programming (LP), which can be effectively solved with some off-the-shelf solving tools, such as CVXPY.

In order to simplify the form of the problem (\ref{problem_uploading_latency_minimization}), we specify $m_i=Kag_i\overline{p}_i-K_i$.
To this end, problem (\ref{problem_uploading_latency_minimization}) can be equivalently transformed into
\begin{subequations}\label{problem_equ_snr_maximization}
	\begin{eqnarray}
		\label{equ_snr_maximization_objective}
		&\min \limits_{ \mathbf{m} } & \frac{\sum_{i=1}^Im_i^2+a^2K^2\sigma^2}{\sum_{i=1}^I(m_i+K_i)^2+a^2K^2\sigma^2} \\
		\label{equ_snr_maximization_m_constraint}
		&{\rm s.t.}&{0\!\le\!m_i\!+\!K_i\!\le\!ag_iK\!\sqrt{p^{
		\rm max}},\ \!\forall i,}\\
		\label{equ_snr_maximization_sum_m_constraint}
		&{}&\sum\nolimits_{i=1}^Im_i^2\le bK^2,
	\end{eqnarray}
\end{subequations}
where $\mathbf{m}=[m_1,\ldots,m_I]^{\rm{T}}$ and constraint (\ref{equ_snr_maximization_m_constraint}) is derived from (\ref{total_latency_minimization_sbs_power_constraint}).
Considering the fractional form of the objective function (\ref{equ_snr_maximization_objective}), we first apply the Dinkelbach parameterization technique to it.
Denote $g(\mathbf{m})=\sum_{i=1}^Im_i^2$ and $h(\mathbf{m})=\tau_{s+1}\sum_{i=1}^I(m_i+K_i)^2$, we can express the optimization problem in the $(s+1)$-th round as
\begin{subequations}\label{problem_1_snr_dc}
	\begin{eqnarray}
		\label{1_snr_dc_objective}
		&\min \limits_{ \mathbf{m} } & g(\mathbf{m})\!-\!h(\mathbf{m})\!+\!\left(\!1\!-\!\tau_{s+1}\!\right)\!a^2\!K^2\!\sigma^2 \\
		\label{1_snr_dc_con}
		&{\rm s.t.}&(\ref{equ_snr_maximization_m_constraint})\ {\rm and}\ (\ref{equ_snr_maximization_sum_m_constraint}),
	\end{eqnarray}
\end{subequations}
where $\tau_{s+1}$ is the Dinkelbach parameter. It can be computed based on 
\begin{equation}\label{tau_update_rule}
		\setlength\abovedisplayskip{2pt}
	    \setlength\belowdisplayskip{2pt}
	    \tau_{s+1}=\frac{\sum_{i=1}^I(m_i^s)^2+a^2K^2\sigma^2}{\sum_{i=1}^I(m_i^s+K_i)^2+a^2K^2\sigma^2},
\end{equation}
where $\mathbf{m}_s=[m_1^s,\ldots,m_I^s]^{\rm{T}}$ is the solution in the $s$-th round.

It can be found that the optimization problem (\ref{problem_1_snr_dc}) is a DC programming, thus the DC Algorithm (DCA) can be used.
The detailed process to solve problem (\ref{problem_equ_snr_maximization}) are summarized in Algorithm \ref{Dinkelbach_DCA_algorithm1}, where $\nabla h\left(\cdot\right)$ is the first order derivative of the function $ h\left(\cdot\right)$.
Note that the convex program in step 7 is actually a convex quadratically constrained quadratic programming (QCQP) problem, which can also be solved efficiently with many off-the-shelf tools. 
Then with convergent output, we can obtain $\overline{p}_i^{*}$ based on $\overline{p}_i^{*}=\left(m_i+K_i\right)/ag_iK\ (i=1,\ldots,I)$.

With $\mathbf{t}$, $\boldsymbol{\rho}$ and $\mathbf{\overline{p}}$, the optimization problem (\ref{reformulated_problem_total_latency_minimization}) becomes
\begin{subequations}\label{problem_equ_mse_minimization}
	\begin{eqnarray}
		\label{equ_mse_minimization_objective}
		&\min \limits_{ \mathbf{c},\mathbf{p} } & \frac{\sum_i\!\left(\!\sum_{k}\!c_{k,i}N_{k,i}\!-\!(\sum_{i}\sum_{k}\!c_{k,i}\!N_{k,i})ag_i\overline{p}_i\!\right)^2}{\left(\sum_{i=1}^I\sum_{k=1}^{N_i}c_{k,i}N_{k,i}\right)^2} \\
		&{\rm s.t.}&(\ref{total_latency_minimization_c_constraint})-(\ref{total_latency_minimization_sensor_power_constraint}),(\ref{total_latency_minimization_mse_constraint}),(\ref{total_latency_minimization_convergence_constraint}),(\ref{re_total_latency_minimization_ct_constraint}),(\ref{re_total_latency_minimization_cct_constraint}).
	\end{eqnarray}
\end{subequations}
For problem (\ref{problem_equ_mse_minimization}), due to the close coupling between $\mathbf{c}$ and $\mathbf{p}$ in constraint (\ref{re_total_latency_minimization_ct_constraint}), it is really difficult to solve it directly.
To tackle this issue, we first solve the sensor selection sub-problem without constraint (\ref{re_total_latency_minimization_ct_constraint}).
Then with the obtained sensor selection scheme $\mathbf{c}$, we manage to find the feasible power control factor of sensors, which satisfies (\ref{total_latency_minimization_sensor_power_constraint}) and (\ref{re_total_latency_minimization_ct_constraint}).
Based on the above analysis, the sensor selection sub-problem to be solved can be expressed as
\begin{subequations}\label{problem_sensor_selection}
	\begin{eqnarray}
		\label{p_sensor_selection_objective}
		&\min \limits_{ \mathbf{c} } & \frac{\sum_i\!\left(\!\sum_{k}\!c_{k,i}N_{k,i}\!-\!(\sum_{i}\sum_{k}\!c_{k,i}\!N_{k,i})ag_i\overline{p}_i\!\right)^2}{\left(\sum_{i=1}^I\sum_{k=1}^{N_i}c_{k,i}N_{k,i}\right)^2} \\
		\label{p_sensor_selection_integer_c_constraint}
		&{\rm s.t.}&c_{k,i}\in\{0,1\},\ \forall k,i,\\
		\label{p_sensor_selection_Ki_constraint}
		&{}& K_i^{\rm min}\le\sum\nolimits_{k=1}^{N_i}c_{k,i}N_{k,i}\le K_i^{\rm max},\ \forall i,\\
		\label{p_sensor_selection_mse_constraint}
		&{}&\!\frac{\!\sum_{i}\!\left(\!\sum_{k}\!c_{k,i}\!N_{k,i}\!-\!(\sum_{i,k}\!c_{k,i}\!N_{k,i})a\!g_i\overline{p}_i\!\right)^2}{\left(\sum_{i=1}^I\!\sum_{k=1}^{N_i}\!c_{k,i}\!N_{k,i}\right)^2}\!\le\! b\!,\\
		\label{p_sensor_selection_convergence_constraint}
		&{}&\!\frac{1+\!\sum_{i=1}^I\left(\sum_{k=1}^{N_i}c_{k,i}N_{k,i}\right)(\rho_i+1)}{\left(\sum_{i,k}c_{k,i}N_{k,i}\right)}\le\frac{\xi}{m},
	\end{eqnarray}
\end{subequations}
where $K_i^{\rm max}=\min\{\frac{\left(T-T_i\right)f_i}{\left(1-\rho_i\right)d^{\rm c}},\sum_{k=1}^{N_i}N_{k,i}\}$.
Because (\ref{p_sensor_selection_objective}) and the left hand of (\ref{p_sensor_selection_mse_constraint}) are in the same form and the initial sensor selection scheme can meet constraint (\ref{p_sensor_selection_mse_constraint}), problem (\ref{problem_sensor_selection}) is always feasible.
To this end, we can remove constraint (\ref{p_sensor_selection_mse_constraint}) from the problem (\ref{problem_sensor_selection}) without affecting its optimal solution and the optimal value of its objective function

Without constraint (\ref{p_sensor_selection_mse_constraint}), problem (\ref{problem_sensor_selection}) is still hard to solve directly because of the non-convexity of formulas (\ref{p_sensor_selection_objective}) and (\ref{p_sensor_selection_convergence_constraint}) and the existence of integer variable $\mathbf{c}$. 
\begin{algorithm}[t!]
	\caption{Algorithm for Solving Problem (\ref{problem_equ_snr_maximization})}
	\label{Dinkelbach_DCA_algorithm1}
	\begin{algorithmic}[1]
		\renewcommand{\algorithmicrequire}{\textbf{Initialize}}
		\renewcommand{\algorithmicensure}{\textbf{Output}}
		\STATE \textbf{Initialize} the tolerance $\delta$, set the iteration number $n_1=0$.
		\STATE Give a feasible vector $\mathbf{m}_0=\{m_1^0,m_2^0,\ldots,m_I^0\}$, which satisfies (\ref{equ_snr_maximization_m_constraint}) and (\ref{equ_snr_maximization_sum_m_constraint}) and compute $\tau_1$ with (\ref{tau_update_rule});
		\REPEAT
		\STATE Set $n_2=0$, $\mathbf{m}_{n_2}^{\prime}=\mathbf{m}_{n_1}$ and $n_{1}\leftarrow n_{1}+1$;
		\REPEAT
		\STATE Set $y_{n_2}=\nabla h\left(\mathbf{m}_{n_2}^{\prime}\right)$;
		\STATE Set $\mathbf{m}_{n_2+1}^{\prime}$ as the optimal solution of the convex program: $\min g\left(\mathbf{m}^{\prime}\right)-\left<\mathbf{m}^{\prime},y_{n_2}\right>$, s.t. (\ref{1_snr_dc_con});
		\STATE Set $n_{2}\leftarrow n_{2}+1$;
		\UNTIL Convergence of $\{\mathbf{m}_{n_2}^{\prime}\}$ and further set $\mathbf{m}_{n_1}\leftarrow \mathbf{m}_{n_2}^{\prime}$;
		\STATE Compute $\tau_{n_1+1}$ with (\ref{tau_update_rule});
		\UNTIL convergence or the maximum iteration is reached;
		\STATE \textbf{Output} the converged solution $\mathbf{m}^{*}$. 
	\end{algorithmic}
\end{algorithm}
To further simplify it, we process it as follows. 
Firstly, the Dinkelbach parameterization technique is applied to deal with the fraction in (\ref{p_sensor_selection_objective}).
Secondly, we replace the integer constraint (\ref{p_sensor_selection_integer_c_constraint}) with the following two formulas:
\begin{equation}\label{p_sensor_selection_re_integer_c_constraint1}
	c_{k,i}\in [0,1],\ \forall k,i,
\end{equation}
\begin{equation}\label{p_sensor_selection_re_integer_c_constraint2}
	\sum\nolimits_{i=1}^I\sum\nolimits_{k=1}^{N_i}c_{k,i}\left(1-c_{k,i}\right)\le 0.
\end{equation}
Thirdly, we make use of a penalty function to further bring the concave constraint (\ref{p_sensor_selection_re_integer_c_constraint2}) into the objective
function.
To this end, in the $(n+1)$-th round of Dinkelbach algorithm iteration, problem (\ref{problem_sensor_selection}) becomes
\begin{subequations}\label{problem_re_sensor_selection}
	\begin{eqnarray}
		\label{p_re_sensor_selection_objective}
		&\min \limits_{ \mathbf{c} } & \sum_{i=1}^I\left(\sum_kc_{k,i}N_{k,i}-\left(\sum_{i,k}c_{k,i}N_{k,i}\right)ag_i\overline{p}_i\right)^2\nonumber \\
		& &-\!\theta_{n+1}\left(\sum_{i,k}c_{k,i}N_{k,i}\right)^2\!+\!\mu\sum_{i,k}\!c_{k,i}\left(1\!-\!c_{k,i}\right) \\
		\label{p_re_sensor_selection_con}
		&{\rm s.t.}&(\ref{p_sensor_selection_Ki_constraint}),(\ref{p_sensor_selection_convergence_constraint})\ {\rm and}\ (\ref{p_sensor_selection_re_integer_c_constraint1}),
	\end{eqnarray}
\end{subequations}
where $\theta_{n+1}$ is the Dinkelbach parameter and $\mu\textgreater 0$ denotes the penalty parameter which penalizes the objective fuction if $c_{k,i}\in\left(0,1\right), (\forall k,i)$.
It is clear that problem (\ref{problem_re_sensor_selection}) is a standard DC programming, which can also be solved by the DCA-based algorithm similar to Algorithm \ref{Dinkelbach_DCA_algorithm1}, except that the sub-problem is replaced by a convex quadratic program (QP).
For the convex QP, some classical algorithms, such as interior-point and active-set methods, can be used. 
Let $\chi\ge 1$ denote the scaling factor of for penalty parameter, $a(\mathbf{c})=\sum_{i=1}^I\left(\sum_kc_{k,i}N_{k,i}-\left(\sum_{i,k}c_{k,i}N_{k,i}\right)ag_i\overline{p}_i\right)^2$ and $b(\mathbf{c})=\theta_{n+1}\left(\sum_{i,k}c_{k,i}N_{k,i}\right)^2\!+\!\mu\sum_{i,k}\!c_{k,i}\left(\!c_{k,i}-1\right)$, we summarize the algorithm for problem (\ref{problem_sensor_selection}) in Algorithm \ref{Dinkelbach_DCA_algorithm2}.

As soon as the solution $\mathbf{c}^{*}$ is obtained, we use greedy rounding to ensure that it strictly meets constraint (\ref{total_latency_minimization_c_constraint}). 
The specific process of greedy rounding is as follows.
\begin{enumerate}
	\item Rounding: round each element in $\mathbf{c}^{*}$ to its nearest integer to obtain $\mathbf{c}_{\rm temp}$.
	\item Computing: substitute $\mathbf{c}_{\rm temp}$ into (\ref{p_sensor_selection_objective}) to obtain $\theta_{\rm temp}$.
	\item Judging: if $\mathbf{c}_{\rm temp}$ satisfies (\ref{p_sensor_selection_Ki_constraint}), (\ref{p_sensor_selection_convergence_constraint}) and $\theta_{\rm temp}\le \theta_1$, then $\mathbf{c}\leftarrow \mathbf{c}_{\rm temp}$, otherwise $\mathbf{c}\leftarrow \mathbf{c}_{0}$.
\end{enumerate}

\begin{algorithm}[t!]
	\caption{Algorithm for Solving Problem (\ref{problem_sensor_selection})}
	\label{Dinkelbach_DCA_algorithm2}
	\begin{algorithmic}[1]
		\renewcommand{\algorithmicrequire}{\textbf{Initialize}}
		\renewcommand{\algorithmicensure}{\textbf{Output}}
		\STATE \textbf{Initialize} the penalty parameter $\mu$, the scaling factor $\chi$ and set the iteration number $n_3=0$.
		\STATE Find a feasible vector $\mathbf{c}_0=\{c_{1,1}^0,c_{2,1}^0,\ldots,c_{N_I,I}^0\}$, which satisfies (\ref{p_sensor_selection_integer_c_constraint})-(\ref{p_sensor_selection_convergence_constraint}) and let $\theta_1=\frac{\sum_i\!\left(\!\sum_{k}\!c_{k,i}^0N_{k,i}\!-\!(\sum_{i}\sum_{k}\!c_{k,i}^0\!N_{k,i})ag_i\overline{p}_i\!\right)^2}{\left(\sum_{i=1}^I\sum_{k=1}^{N_i}c_{k,i}^0N_{k,i}\right)^2}$;
		\REPEAT
		\STATE Set $n_4=0$, $\mathbf{c}_{n_4}^{\prime}=\mathbf{c}_{n_3}$ and $n_{3}\leftarrow n_{3}+1$;
		\REPEAT
		\STATE Set $y_{n_4}=\nabla b\left(\mathbf{c}_{n_4}^{\prime}\right)$;
		\STATE Set $\mathbf{c}_{n_4+1}^{\prime}$ as the optimal solution of the convex program: $\min a\left(\mathbf{c}^{\prime}\right)-\left<\mathbf{c}^{\prime},y_{n_2}\right>$, s.t.(\ref{p_re_sensor_selection_con});
		\STATE Set $n_{4}\leftarrow n_{4}+1$;
		\UNTIL Convergence of $\{\mathbf{c}_{n_2}^{\prime}\}$ and further set $\mathbf{c}_{n_3}\leftarrow \mathbf{c}_{n_4}^{\prime}$;
		\STATE Substituting $\mathbf{c}_{n_3}$ into (\ref{p_sensor_selection_objective}) and assigning the resulting value to $\theta_{n_3+1}$, $\mu\leftarrow \chi\mu$ ;
		\UNTIL convergence or the maximum iteration is reached;
		\STATE \textbf{Output} the optimal or the converged solution $\mathbf{c}^{*}$. 
	\end{algorithmic}
\end{algorithm}

\begin{algorithm}[t!]
	\caption{Overall Algorithm}
	\label{overall_algorithm}
	\begin{algorithmic}[1]
		\renewcommand{\algorithmicrequire}{\textbf{Initialize}}
		\renewcommand{\algorithmicensure}{\textbf{Output}}
		\STATE \textbf{Initialize} the sensor selection scheme $\mathbf{c}$ and the vector of power control factor $\mathbf{p}$.
		\REPEAT
		\STATE Given $\mathbf{c}$ and $\mathbf{p}$, solve problem (\ref{problem_T_minimization}) to obtain $\mathbf{t}$ and $\mathbb{\rho}$;
		\STATE Given fixed $\mathbf{c}$ and $\mathbf{p}$, solve $\mathbf{m}$ with Algorithm \ref{Dinkelbach_DCA_algorithm1} and then compute $\mathbf{\overline{p}}$;
		\STATE Given fixed $\mathbf{t}$, $\mathbb{\rho}$ and $\mathbf{\overline{p}}$, compute $\mathbf{c}^*$ with Algorithm \ref{Dinkelbach_DCA_algorithm2};
		\STATE Update $\mathbf{c}^*$ with greedy rounding;
		\STATE Find one feasible power control factor solution $\mathbf{p}$;
		\UNTIL convergence;
		\STATE \textbf{Output} the optimal solution $\{\mathbf{t}^*,\mathbb{\rho}^*,\mathbf{c}^*,\mathbf{p}^*,\mathbf{\overline{p}}^{*}\}$. 
	\end{algorithmic}
\end{algorithm}

This step is not always necessary but related to the value of $\mu$. The specific relationship will be given after the overall algorithm is presented.
With the obtained $\mathbf{c}^{*}$, we manage to find one feasible power control factor solution $\mathbf{p}$ that satisfies constraints (\ref{total_latency_minimization_sensor_power_constraint}) and (\ref{re_total_latency_minimization_ct_constraint}).
For the selected sensors at the same SBS, the achievable transmission rate of the last decoded sensor does not contain any interference.
Therefore, we can solve the power control factor of the selected sensors in the reverse order of the decoding order according to $T_i$ in (\ref{re_total_latency_minimization_ct_constraint}).
When all calculated power control factor of sensors meet (\ref{total_latency_minimization_sensor_power_constraint}), then one feasible power control factor $\mathbf{p}$ is obtained. 

According to above discussions, the detailed process of solving original problem (\ref{reformulated_problem_total_latency_minimization}) is given in Algorithm \ref{overall_algorithm}.
Note that the value of $\mu$ in sub-problem (\ref{problem_re_sensor_selection}) controls the extent to which the output $\mathbf{c}^{*}$ of Algorithm \ref{Dinkelbach_DCA_algorithm2} violates constraint (\ref{p_sensor_selection_integer_c_constraint}).
Specifically, with sufficient large $\mu$, $\mathbf{c}^{*}$ will strictly meet the constraint (\ref{p_sensor_selection_integer_c_constraint}).
To this end, the sixth step in Algorithm \ref{overall_algorithm} will not cause any performance deterioration.
As for the complexity and convergence of Algorithm 3, they are given below:

\textit{Complexity:} For the convex optimization sub-problems in Algorithm \ref{Dinkelbach_DCA_algorithm1} and \ref{Dinkelbach_DCA_algorithm2}, the interior point method is considered to solve them.
The overall algorithm shown in Algorithm \ref{overall_algorithm} is an iterative algorithm, and the complexity of executing it for one iteration is $\mathcal{O}\left(\left(2I+1\right)^2\left(2I+\!\sum_{i=1}^IN_i+1\right)\!+\!T_1T_2I^3\notag\right. \\ \left.+T_3T_4\left(\!\sum_{i=1}^IN_i\right)^3+2\sum_{i=1}^IN_i+I\right)$.
Specifically, the complexity of solving linear programming problem in the third line of Algorithm \ref{overall_algorithm} is $\mathcal{O}\left(\left(2I+1\right)^2\left(2I+\!\sum_{i=1}^IN_i+1\right)\right)$.
The complexity of Algorithm \ref{Dinkelbach_DCA_algorithm1} is $\mathcal{O}\left(T_1T_2I^3\right)$, where $T_1$ and $T_2$ are the iteration number of its outer and inner layer, respectively.
And the complexity of calculating $\mathbf{\overline{p}}$ with $\mathbf{m}$ is $\mathcal{O}\left(I\right)$.
As for Algorithm \ref{Dinkelbach_DCA_algorithm2}, its complexity is $\mathcal{O}\left(T_3T_4\left(\!\sum_{i=1}^IN_i\right)^3\right)$, where $T_3$ and $T_4$ denote the iteration number of its outer and inner layer, respectively.
Finally, the computational complexity of the sixth and seventh lines in Algorithm \ref{overall_algorithm} are both $\mathcal{O}\left(\sum_{i=1}^IN_i\right)$.
Since the computational complexity of greedy rounding is low and that of the overall algorithm is at polynomial level, Algorithm \ref{overall_algorithm} is applicable to the scenarios with a large number of SBSs and sensors.

\textit{Convergence:} 
Let $(\mathbf{t}^{(l)},\boldsymbol{\rho}^{(l)}, \mathbf{c}^{(l)}, \mathbf{p}^{(l)},\mathbf{\overline{p}}^{(l)})$ and function $H(\mathbf{t}^{(l)},\boldsymbol{\rho}^{(l)}, \mathbf{c}^{(l)}, \mathbf{p}^{(l)},\mathbf{\overline{p}}^{(l)})$ denote the solution in the $l$-th iteration of Algorithm 3 and the value of (\ref{re_total_latency_minimization_objective}), respectively.
Problem (\ref{problem_T_minimization}) is a linear programming problem for which the optimal solution is obtained.
Algorithm \ref{Dinkelbach_DCA_algorithm1} and \ref{Dinkelbach_DCA_algorithm2} are both based on the Dinkelbach parameterization technique and DC programming.
According to \cite{dinkelbach1967nonlinear} and the fact that the value of the new objective function in the DC programming is not less than the original objective function at any feasible  point, it can be obtained that $\theta$ and $\tau$ are monotonically non-increasing during the iteration process of Algorithm \ref{Dinkelbach_DCA_algorithm1} and \ref{Dinkelbach_DCA_algorithm2}.
Finally, for the sixth and seventh lines of Algorithm \ref{overall_algorithm}, there is a judgment that the objective function value does not increase during the rounding process of sensor selection variables, and the corresponding sensor power control factor can be calculated in a closed form.
To this end, it can be found that $H(\mathbf{t}^{(l+1)},\boldsymbol{\rho}^{(l+1)}, \mathbf{c}^{(l+1)}, \mathbf{p}^{(l+1)},\mathbf{\overline{p}}^{(l+1)})\le H(\mathbf{t}^{(l)},\boldsymbol{\rho}^{(l)}, \mathbf{c}^{(l)}, \mathbf{p}^{(l)},\mathbf{\overline{p}}^{(l)})$.
Overall, during the iteration process of Algorithm \ref{overall_algorithm}, the objective function value is monotonic non-increasing and lower-bounded by zero, so it converges.

\section{Numerical Results}
In the simulation, we consider an IoT network consisting of one MBS and five SBSs. 
The global FL model is a multilayer perceptron (MLP) with two hidden layers of 200 and 100 neurons, respectively.
Cross-entropy is used as its loss function and learning rate is set to 0.3.
The learning tasks are handwritten digit recognition on the MNIST dataset and fashion products classification on the Fashion-MNIST dataset.
Both datasets contain 70,000 images from 10 categories, including 60,000 images for training and 10,000 images for testing.
In our simulation, the datasets collected by different SBSs from sensors within their range are Non-IID, that is, the main categories of the datasets are different.
The SBSs are distributed within a range of 20 to 100 meters from the MBS.
There are three sensors served by each SBS, with a distance of [10, 80] meters from corresponding SBS.
The post-transmission factor is set to 4.
In each round, each sensor can send 20 samples with each data size of 0.1 Mbit to its serving SBS for the local training.
The maximum transmit power of sensors and SBSs are set to 0.2 W and 4 W, respectively.  
Other more detailed simulation parameters are shown in Table \ref{paramater_table}.

To verify the performance of the proposed algorithm, we set the following schemes as benchmarks.
a) All sensor selection: In each round all sensors transmit local data samples to the SBSs, 
b) Random sensor selection: In each round, each SBS randomly selects sensors that meet the number of samples required for its local training,
c) Fixed pruning rate 0.1: In each round, the pruning rates of FL model in SBSs are set to fixed values of 0.1,
d) Ideal FL algorithm: In each round, each SBS selects all sensors and the local model pruning rate of each SBS is 0.
The coefficient of gradient aggregation is strictly in accordance with the proportion of the sample number used in local training to the total sample and the noise during over-the-air computation is neglected,
e) Perfect aggregation algorithm: In each round, this scheme maintains the same pruning rates and sensor selection as the proposed algorithm with $\xi=140$, but the gradient aggregation process is the same as above ideal FL algorithm. 
In the following part, we present the performance of the proposed algorithm in terms of the one-round FL latency and recognition accuracy.

\begin{table}[t!]
	\setlength{\abovecaptionskip}{-0.3mm}
	\caption{Parameter Settings}\label{paramater_table}
	\centering
	\begin {tabular}{|c|c|c|c|}
	\hline
	\textbf{Parameter}&\textbf{Value}&\textbf{Parameter}&\textbf{Value}\\
	\hline
	$D_{\rm M}$&10 Mbit&$B$&5 MHz\\
	\hline
	$d^{\rm c}$&0.168 GHz&$f_i\ (\forall i)$&10 GHz\\
	\hline
	$K_i^{\rm min}\ (\forall i)$&40&$B_{\rm M}$&3 MHz\\
	\hline
	$\rho_i^{\rm min}\ (\forall i)$&0.1&$\rho_i^{\rm max}\ (\forall i)$&0.7 \\
	\hline
	$\mu$&30&$\chi$&1 \\
	\hline
\end{tabular}
\end{table}

\begin{figure} [t!]
	\centering
	\includegraphics[height=2.4 in,width=3.5 in]{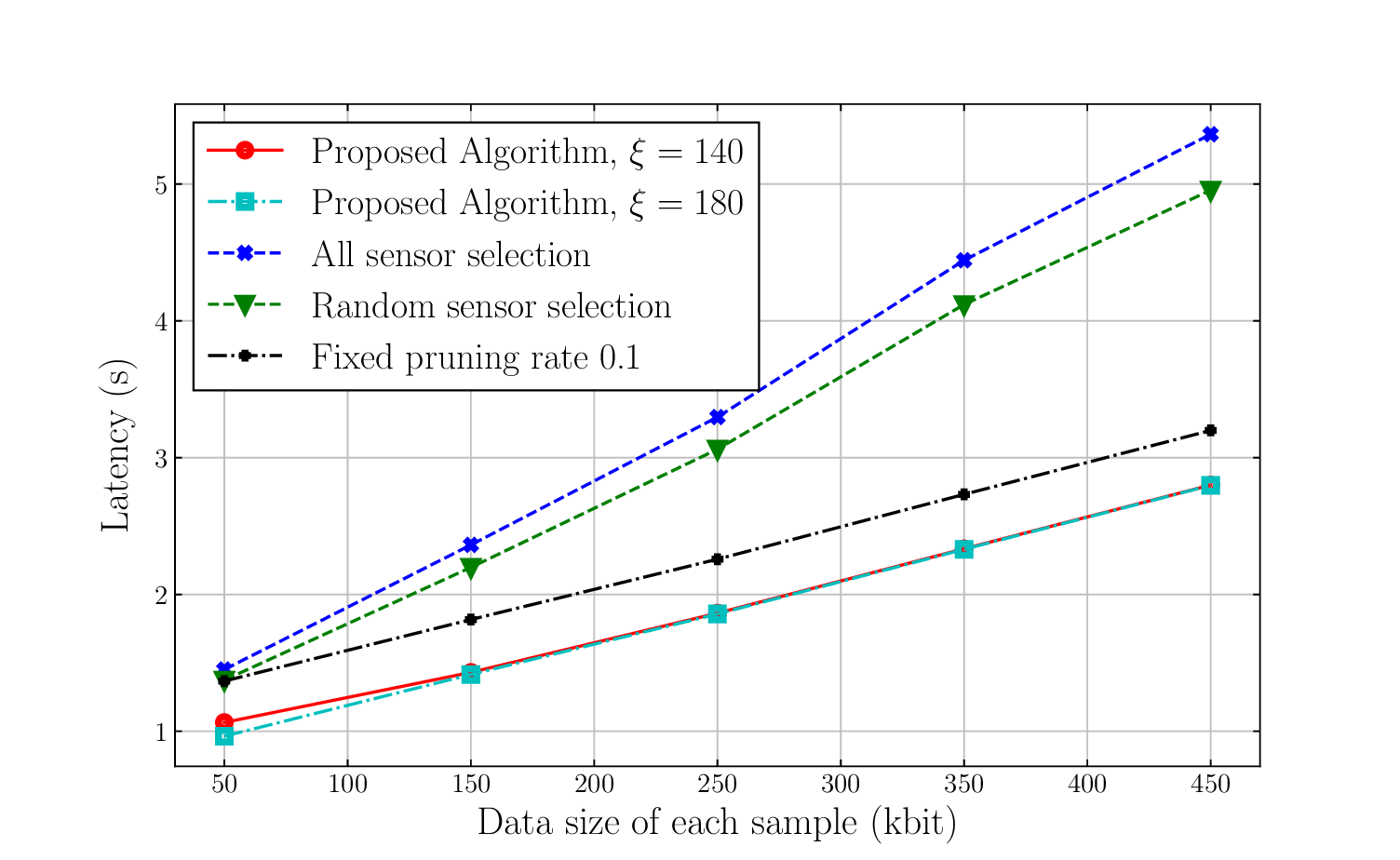}
	\vspace{-3mm}
	\caption{Latency vs. data size of each sample}
	\label{sample_data_size}	
\end{figure}

\begin{figure} [t!]
	\centering
	\includegraphics[height=2.4 in,width=3.5 in]{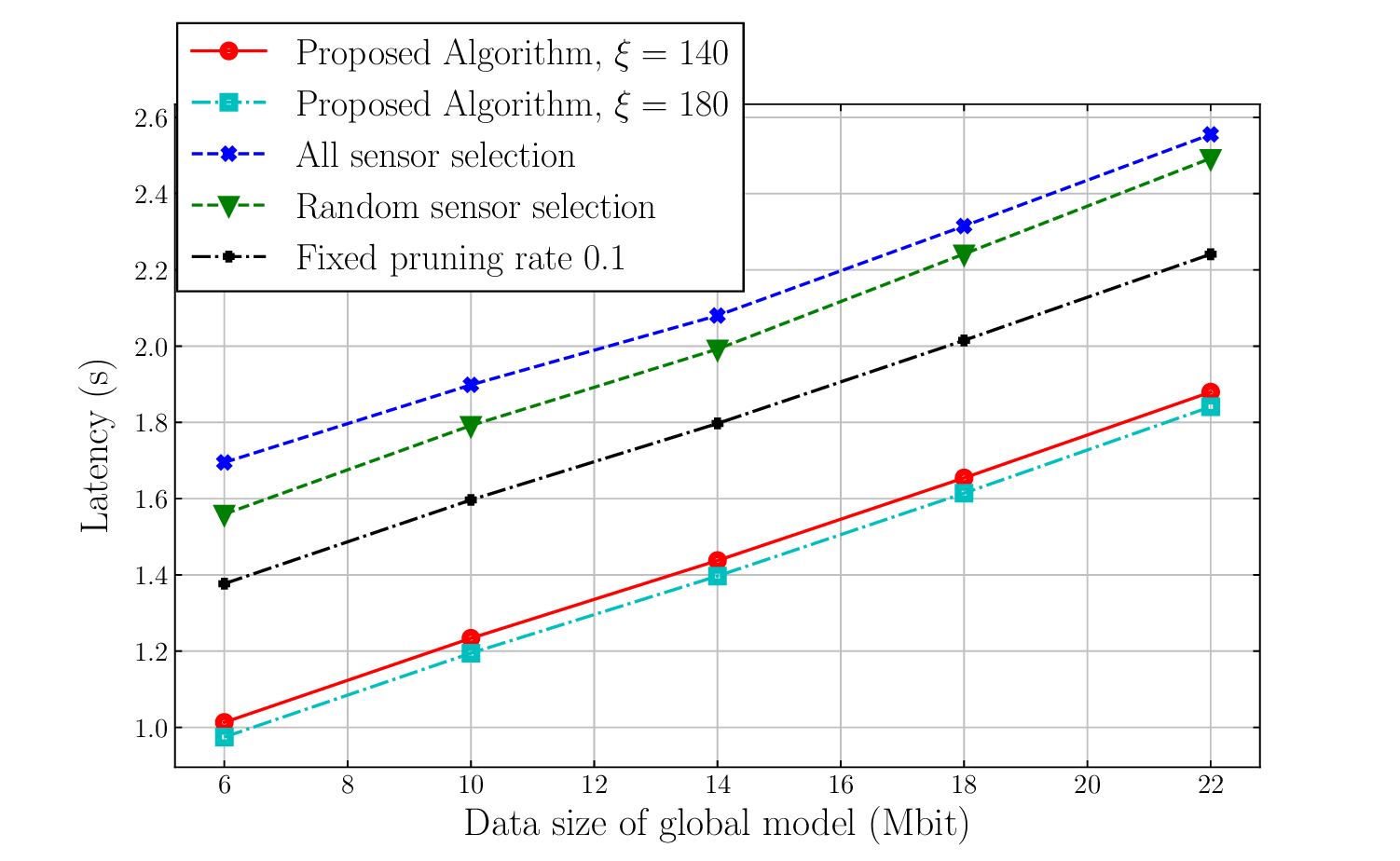}
	\vspace{-3mm}
	\caption{Latency vs. data size of global model}
	\label{model_data_size}
\end{figure}

\begin{figure} [t!]
	\centering
	\includegraphics[height=2.4 in,width=3.5 in]{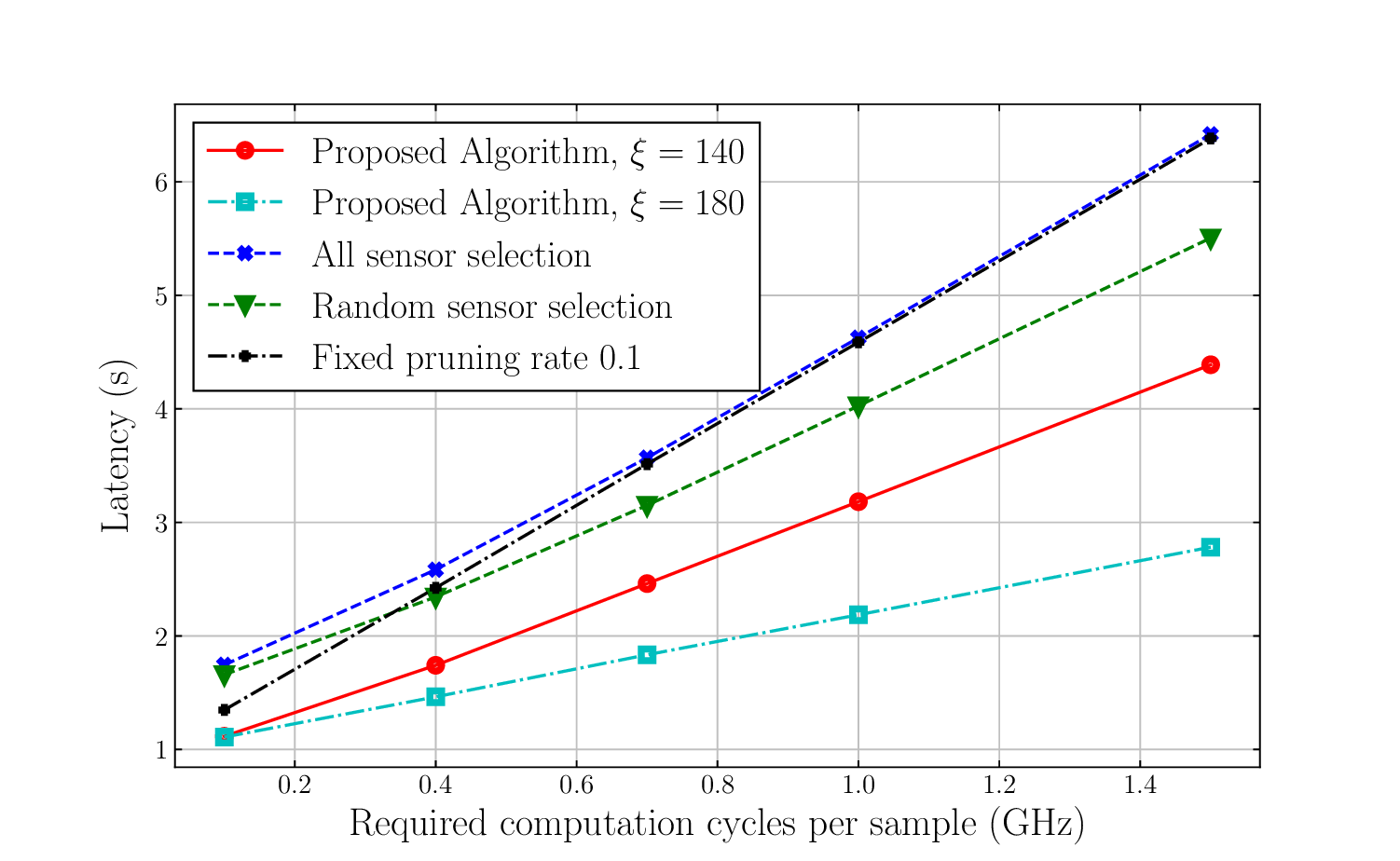}
	\vspace{-3mm}
	\caption{Latency vs. required computation cycles per sample}
	\label{comp_burden}
\end{figure}

\begin{figure} [t!]
	\centering
	\includegraphics[height=2.4 in,width=3.5 in]{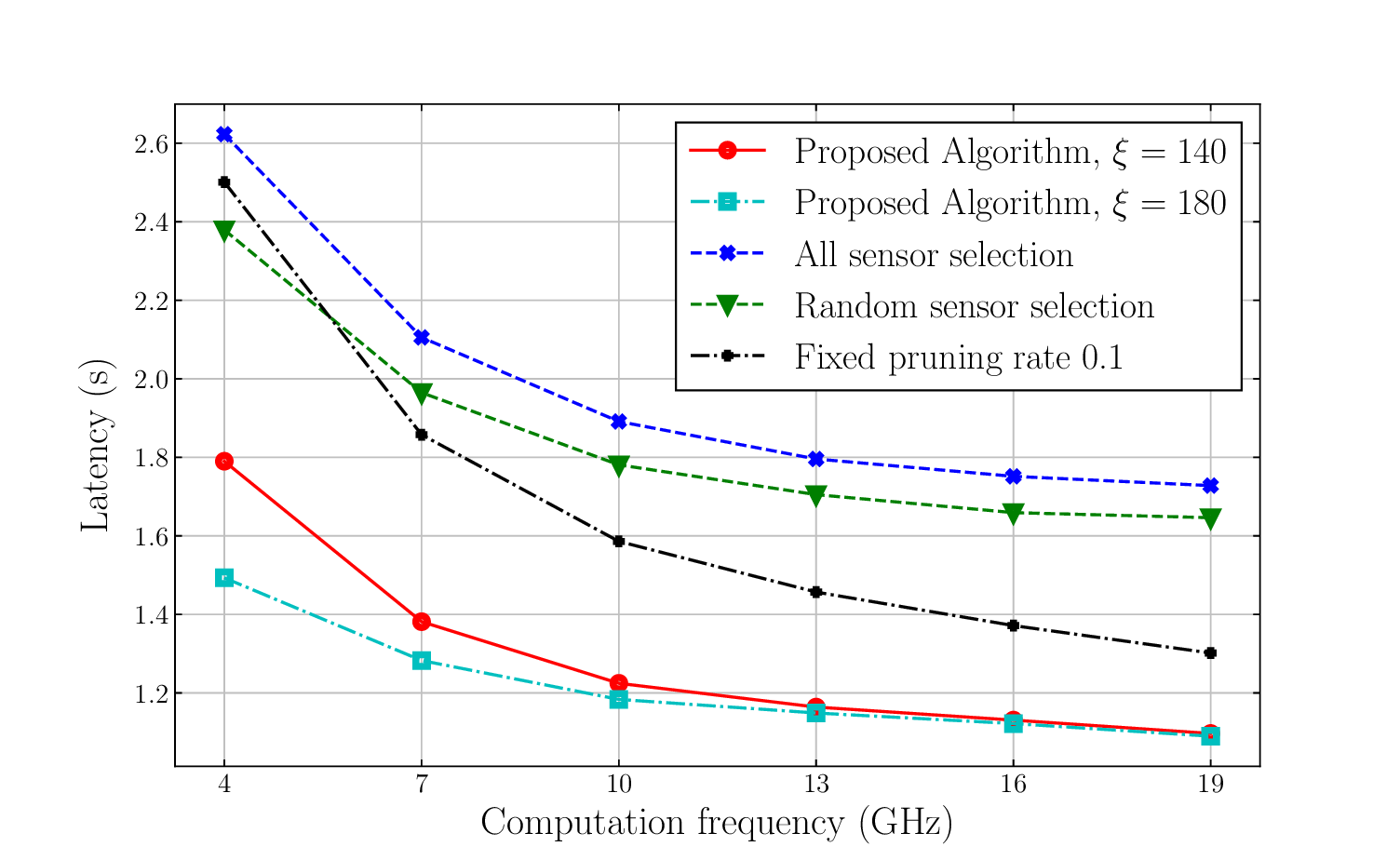}
	\vspace{-3mm}
	\caption{Latency vs. computation frequency of SBS}
	\label{comp_frequency}
\end{figure}

In Fig. \ref{sample_data_size} and Fig. \ref{model_data_size}, we show the effect of the data size of each data sample and global model on the latency, respectively.
To be specific, it can be seen from Fig. \ref{sample_data_size} that FL latency increases with the increase of the data size of each sample.
The fixed pruning rate algorithm always brings higher latency than proposed algorithms with $\xi=140$ and $\xi=180$, because it cannot adaptively adjust the local pruning rates.
In addition, for all sensor selection algorithm, the latency will be greatly deteriorated due to the high interference when sensors uploading their data samples to SBSs with NOMA.
With larger $\xi$, the looser the convergence constraint (\ref{total_latency_minimization_convergence_constraint}) is, so the proposed algorithm can use higher pruning rates and more flexible sensor selection schemes to further reduce the FL latency. 
However, when the data size of each sample increases, the difference in sample collection latency among SBSs increases, the latency reduction caused by increasing $\xi$ of the proposed algorithm becomes insignificant.
In Fig. \ref{model_data_size}, with the increase of the data size of FL model, the latency achieved by all algorithms increase approximately linearly. 
For random sensor selection algorithm, it usually has a lower computation rate due to its larger signal MSE, thus its latency is more significantly affected by the data size of model.
However,under all settings of the data size of global model, the proposed algorithm with $\xi=140$ and $\xi=180$ always achieves lower FL latency than benchmarks.

In Fig. \ref{comp_burden}, the effect of the required computation cycles per sample on latency is presented.
It can be seen that the latency increases with the increase of the required computation cycles for all algorithms.
Note that the growth trend of the proposed algorithm with $\xi=140$ and $\xi=180$, all sensor selection algorithm and random sensor selection algorithm is not completely linear, but gradually close to linear.
With the increase of $d_c$, the impact of data collection latency on one-round FL latency decreases gradually, so the difference on SBSs' local pruning decreases and the pruning rates will make the constraint (\ref{total_latency_minimization_convergence_constraint}) equation hold.
Therefore, when $d_c$ is large enough, the slope of the curve of the above algorithms is almost constant and very close to linear.
By comparing the proposed algorithm with $\xi=140$ and $\xi=180$, we can find that with the increase of required computation cycles, larger $\xi$ brings more benefits to reducing latency, because it allows higher local pruning rates.
With the increase of the required computation cycles per sample, the effect of local training latency on total latency increases.
To this end, the latency of algorithm with fixed pruning rate 0.1 increases rapidly.
Note that with any $d_c$, the latency of the proposed algorithms with $\xi=140$ and $\xi=180$ are lower than other schemes, which confirms the performance of proposed algorithm.

The effect of the computation frequency of each SBS is further studied in Fig. \ref{comp_frequency}.
With the increase of computation frequency, the latency of all schemes shows a downward trend.
When the computation frequency of SBSs are large enough, the local training latency has a much smaller impact on latency of SemiFL than sample transmission latency and global aggregation latency.
Therefore, the latency will not be significantly affected with the further increase of computation frequencies of SBSs.
By comparing the proposed algorithm with $\xi=140$ and $\xi=180$, it can be seen that the increase of computation frequency of each SBS, the latency improvement brought by increasing $\xi$ also decreases.

\begin{figure} [t!]
	\centering
	\includegraphics[height=2.4 in,width=3.5 in]{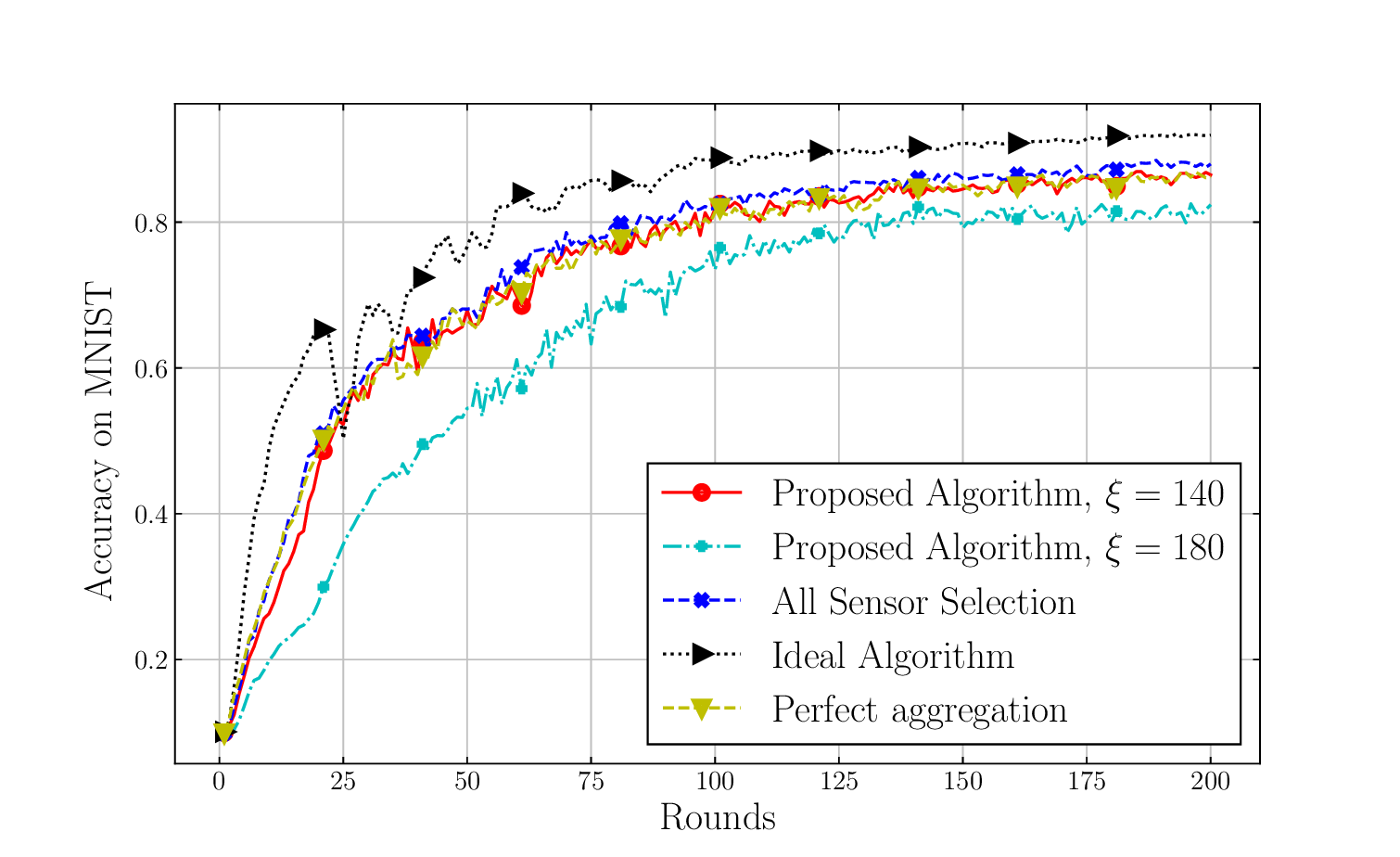}
	\vspace{-3mm}
	\caption{Identification accuracy on MNIST dataset}
	\label{accuracy_mnist}
\end{figure}

\begin{figure} [t!]
	\centering
	\includegraphics[height=2.4 in,width=3.5 in]{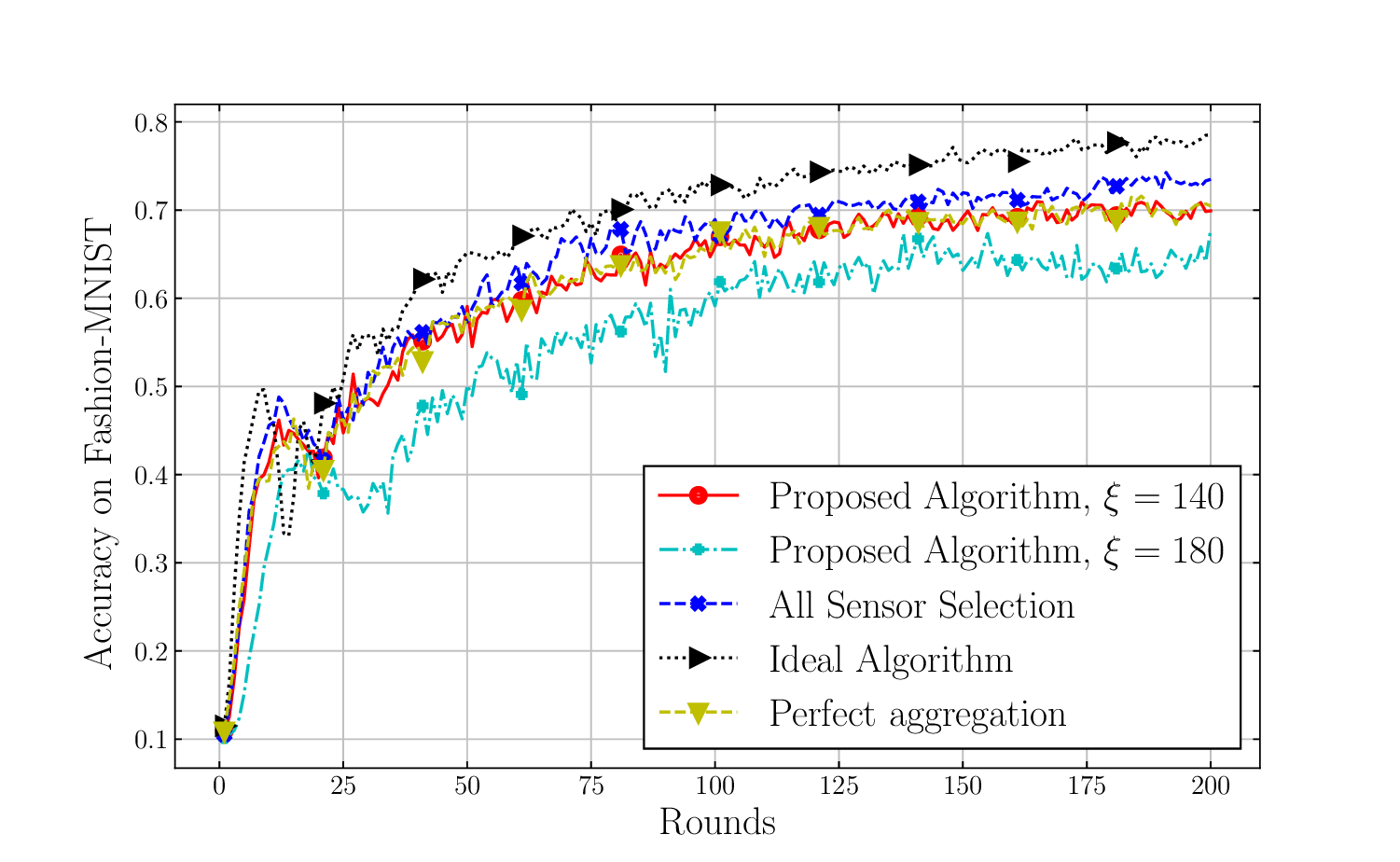}
	\vspace{-3mm}
	\caption{Identification accuracy on Fashion-MNIST dataset}
	\label{accuracy_f_mnist}
\end{figure}

In Fig. \ref{accuracy_mnist}, the simulation result on MNIST dataset is shown.
The ideal FL algorithm always has the best learning performance because there is no influence of pruning and noise, and the aggregation coefficients equal to the proportion of the local sample number at SBSs to the total number.
Then, the accuracy of all sensor selection algorithm, perfect aggregation algorithm and the proposed algorithm with $\xi=140$ is reduced compared with ideal FL algorithm, because they use higher local pruning rates to reduce FL latency.
Despite the existence of aggregation coefficient errors and noise, the accuracy of the proposed algorithm with $\xi=140$ is always close to that of the perfect aggregation algorithm, which proves the effectiveness of the proposed algorithm in controlling SBSs' power control factor.
Besides, the accuracy of all sensor selection algorithm is always slight higher than that of the proposed algorithm with $\xi=140$.
Although it cannot flexibly adjust the sensor selection scheme, using all samples for training guarantees its learning performance.
When $\xi=180$, the constraint of the single-round convergence upper bound (\ref{total_latency_minimization_convergence_constraint}) is loose enough that it hardly works, so the identification accuracy of proposed algorithm with $\xi=180$ is much lower than others.   

In Fig. \ref{accuracy_f_mnist}, the identification accuracy on Fashion-MNIST dataset of different algorithms is given.
Similar to the result in Fig. \ref{accuracy_mnist}, the identification accuracy of the ideal FL algorithm is the highest among all algorithms.
Affected by the more flexible pruning rates or sensor selection scheme for reducing FL latency, the recognition accuracy of the other for schemes is lower than ideal FL algorithm, but the gap is small on Fashion-MNIST dataset.
Note that the recognition accuracy of the proposed algorithm with $\xi=140$ is always close to the accuracy of all sensor selection algorithm and the perfect aggregation algorithm, which proves its performance.

\section{Conclusions and Future Work}
In this paper, we proposed a SemiFL framework, where each SBS collected data samples for distributed training, and then their local gradients were aggregated at MBS.
To help reduce its training and communication latency, the network pruning and over-the-air computation were jointly applied to this framework.
We first analyzed the convergence upper bound of the proposed SemiFL, which was affected by many factors, such as data heterogeneity, power control mechanism, wireless channel noise and network pruning rates.
Then a convergence-constrained FL latency minimization problem by jointly optimizing various wireless resources and pruning rates was formulated.
For this complex optimization problem, we decoupled it into several sub-problems and designed corresponding algorithms respectively.
Finally, the numerical simulations were conducted to demonstrate the effectiveness of our proposed algorithm.

With the increasing complexity of learning tasks, neural networks with more complex structures, such as residual and recurrent neural network, are gradually being widely used.
For these complex neural networks, pruning different neurons may have different effects on the training latency and learning performance.
Therefore, further detailed analysis of the above modeling process is needed, which is one of our future directions.
In addition, during the data collection of SBSs, sensors upload samples through wireless channels.
Thus, the privacy protection scheme of this process is worth further research. 
Finally, there may be malicious nodes intentionally disrupting model training in actual networks.
In this regard, it is also of great research value to design the corresponding scheme that can recognize and resist them.

\appendices
\section{Proof of Theorem 1}
Based on (\ref{global_model_update_rule}), we rewrite the global model update rule as
\begin{equation}\tag{A-1}
	W^{t+1}=W^t-\eta\left(\nabla{F\left({W}^t\right)}-o^t\right),
\end{equation}
where $o^t=\nabla{F\left({W}^t\right)}-\hat{G}^t$.
Since $F(\cdot)$ is $\beta$-smooth, we have
\begin{equation}\label{A-2}\tag{A-2}
	\begin{aligned}
	&F\left(W^{t+1}\right)\\&\le\! F\left(W^t\right)\!+\!\left<W^{t+1}-W^t,\nabla{F\left({W}^t\right)}\right>\!+\!\frac{\beta}{2}\left\|W^{t+1}-W^t\right\|^2\\&\!=\!F\left(\!W^t\!\right)\!-\!\eta\left<\!\nabla{F\left({W}^t\right)}\!-\!o^t,\nabla{F\left({W}^t\right)}\!\right>\!+\!\frac{\left\|\nabla{F\left(\!{W}^t\!\right)}\!-\!o^t\right\|^2}{2/(\beta\eta^2)}.
	\end{aligned}
\end{equation}
Given $\eta=\frac{1}{\beta}$, we can get
\begin{equation}\label{A-3}\tag{A-3}
	\begin{aligned}
		F\left(W^{t+1}\right)&\le F\left(W^t\right)-\frac{1}{\beta}\left<\nabla{F\left({W}^t\right)},\nabla{F\left({W}^t\right)}\right>\\&\ \ \ +\frac{1}{2\beta}\left\|\nabla{F\left({W}^t\right)}\right\|^2+\frac{1}{2\beta}\|o^t\|^2\\&=\!F\left(W^t\right)\!-\!\frac{1}{2\beta}\left\|\nabla{F\left({W}^t\right)}\right\|^2\!+\!\frac{1}{2\beta}\|o^t\|^2.
	\end{aligned}
\end{equation}

Next we focus on derive the expression of $\mathbb{E}[\|o^t\|^2]$, as
\begin{equation}\label{A-4}\tag{A-4}
	\begin{aligned}
		\mathbb{E}[\|o^t\|^2]&= \mathbb{E}\Bigg[\left\|\nabla{F\left({W}^t\right)}-\frac{1}{K^t}\sum_{i=1}^IK_i^tG_i^{t}+\frac{1}{K^t}\sum_{i=1}^IK_i^tG_i^{\rm t}\right.\\&\left.\ \ \ \ \ \ \ \ \ -\sum_{i=1}^Iag_i^t{\overline{p}_i^t}G_i^{t}-a\sigma_{\rm n}^t\mathbf{z}_0 \right\|^2\Bigg]\\&\le\!2\mathbb{E}\!\left[\left\|\nabla{F\left({W}^t\right)}\!-\!\frac{1}{K^t}\sum_{i=1}^IK_i^tG_i^{t}\right\|^2\right]\!+\\&\ \ \ \ \!2\mathbb{E}\!\left[\left\|\frac{1}{K^t}\sum_{i=1}^IK_i^tG_i^{t}\!-\!\sum_{i=1}^Iag_i^t{\overline{p}_i^t}G_i^{t}\!-\!a\sigma_{\rm n}^t\mathbf{z}_0\right\|^2\right]\\&=2\mathbb{E}\left[\|\epsilon^t\|^2\right]+2Q{a}^2(\sigma_{\rm n}^t)^2\sigma^2\\&\ \ \ \ +\frac{2}{(K^t)^2}\mathbb{E}\left[\left\|\sum_{i=1}^I\left(K_i^t-K^tag_i^t{\overline{p}_i^t}\right)G_i^{t}\right\|^2\right]\\
		&\le\frac{2I}{(K^t)^2} \sum_{i=1}^I\left\{\left(K_i^t-Kag_i^t{\overline{p}_i^t}\right)^2\mathbb{E}\left[\left\|G_i^{t}\right\|^2\right]\right\}\\&\ \ \ \ +2\mathbb{E}\left[\|\epsilon^t\|^2\right]+2Q{a}^2(\sigma_{\rm n}^t)^2\sigma^2 \\&\le\frac{2IM^2}{(K^t)^2} \sum_{i=1}^I\left(K_i^t-K^tag_i^t{\overline{p}_i^t}\right)^2\\&\ \ \ \ +2\mathbb{E}\left[\|\epsilon^t\|^2\right]+2Q{a}^2(\sigma_{\rm n}^t)^2\sigma^2,
	\end{aligned}
\end{equation}
where $\epsilon^t=\nabla{F\left({W}^t\right)}-\frac{1}{K^t}\sum_{i=1}^IK_i^tG_i^{t}$.
Note taht the second item is related to the MSE of the signal. 
Considering that MSE has an important impact on both aggregation rate and convergence behavior, constraint $\frac{1}{(K^t)^2} \sum_{i=1}^I\left(K_i^t-K^tag_i^t{\overline{p}_i^t}\right)^2\le b$ is introduced.
Besides, since the dimension of local gradients $Q$ is obviously  finite, $\sigma_{\rm n}^t$ is upper bounded \cite{zou2023knowledge,wang2022federated}.  
Let ${a}^2(\sigma_{\rm n}^t)^2\le {a^{\prime}}^2$, (\ref{A-4}) can be reformulated as
\begin{equation}\label{A-5}\tag{A-5}
	\mathbb{E}\left[\|o^t\|^2\right]\le 2bIM^2+2\mathbb{E}\left[\|\epsilon^t\|^2\right]+2Q{a^{\prime}}^2\sigma^2.
\end{equation}

As for $\mathbb{E}[\|\epsilon^t\|^2]$, the following inequality is derived:
\begin{equation}\notag{}
	\begin{aligned}
	&\mathbb{E}\left[\|\epsilon^t\|^2\right]=\mathbb{E}\left[\left\|\nabla{F\left({W}^t\right)}-\frac{1}{K^t}\sum_{i=1}^IK_i^tG_i^{\rm t}\right\|^2\right]\\&=\!\mathbb{E}\!\Bigg[\!\left\|\nabla{F\left({W}^t\right)}\!-\!\frac{\sum_{i}K_i^t\nabla{F_i\left({W}^t\right)}}{K^t}\!+\!\frac{\sum_{i}K_i^t\nabla{F_i\left({W}^t\right)}}{K^t}\!\right.\\&\left.\ \ \ \ \   \!-\!\frac{\sum_{i}\!K_i^t\nabla{F_i\left(\tilde{W}_i^t\right)}}{K^t}  \!+\!\frac{\sum_{i}\!K_i^t\nabla{F_i\left(\tilde{W}_i^t\right)}}{K^t}\!-\!\frac{\sum_{i}K_i^tG_i^{t}}{K^t}\!\right\|^2\!\Bigg]\\
	&\le\!\frac{3}{(K^t)^2}\!\mathbb{E}\!\left[\!\left\|\!\sum_{i=1}^I\!K_i^t\!\left(\nabla{F\left({W}^t\right)}\!-\!\nabla{F_i\left({W}^t\right)}\right)\!\right\|^2\!\right]\\&\ \  +\!\frac{3}{(K^t)^2}\!\mathbb{E}\!\left[\!\left\|\sum_{i=1}^IK_i^t\left(\nabla{F_i\left({W}^t\right)}\!-\!\nabla{F_i\left(\tilde{W}_i^t\right)}\right)\!\right\|^2\!\right]\\&\ \  +\frac{3}{(K^t)^2}\mathbb{E}\left[\left\|\sum_{i=1}^IK_i^t\left(\nabla{F_i\left(\tilde{W}_i^t\right)}-G_i^{\rm t}\right)\right\|^2\right]
	\end{aligned}
\end{equation}
\begin{equation}\tag{A-6}\label{A-6}
	\begin{aligned}
	&\le\!\frac{3I}{(K^t)^2}\!\sum_{i=1}^I(K_i^t)^2\!\mathbb{E}\!\left[\!\left\|\nabla{F\left({W}^t\right)}\!-\!\nabla{F_i\left({W}^t\right)}\right\|^2\!\right]\\&\ \ +\!\frac{3I}{(K^t)^2}\!\sum_{i=1}^I(K_i^t)^2\!\mathbb{E}\!\left[\!\left\|\nabla{F_i\left({W}^t\right)}\!-\!\nabla{F_i\left(\tilde{W}_i^t\right)}\right\|^2\!\right]\\&\ \  +\frac{3I}{(K^t)^2}\sum_{i=1}^I(K_i^t)^2\mathbb{E}\left[\left\|\nabla{F_i\left(\tilde{W}_i^t\right)}-G_i^{\rm t}\right\|^2\right]\\&\overset{(a)}{\le}\!\frac{3IU^2}{(K^t)^2}\!\sum_{i=1}^I(K_i^t)^2\!+\!\frac{3I\beta^2}{(K^t)^2}\!\sum_{i=1}^I(K_i^t)^2\mathbb{E}\!\left[\!\left\|{W}^t-\tilde{W}_i^t\right\|^2\!\right] \\&\ \ +\!\frac{3I}{(K^t)^2}\!\sum_{i=1}^I(K_i^t)^2\mathbb{E}\!\left[\!\left\|\nabla{F_i\left(\tilde{W}_i^t\right)}\!-\!G_i^{\rm t}\right\|^2\!\right]\\&\overset{(b)}{\le}\frac{3IU^2}{(K^t)^2}\sum_{i=1}^I(K_i^t)^2+\frac{3I\beta^2D^2}{(K^t)^2}\sum_{i=1}^I(K_i^t)^2\rho_i^t \\&\ \ +\frac{3I}{(K^t)^2}\sum_{i=1}^I(K_i^t)^2\mathbb{E}\left[\left\|\nabla{F_i\left(\tilde{W}_i^t\right)}-G_i^{\rm t}\right\|^2\right],
	\end{aligned}
\end{equation}
where step (a) is based on Assumption 4 and $\beta$-smooth ($\|\nabla{F_i\left(W^t\right)}-\nabla{F_i(\tilde{W}_i^t)}\|\le\beta\|W^t-\tilde{W}_i^t\|$) and step (b) stems from $\mathbb{E}[\|W^t-\tilde{W}_i^t\|^2]\le\rho_i^tE[\|W^t\|^2]\le\rho_i^tD^2$.

As analyzed in \cite{zeng2022wiresly,mccandlish2018empirical}, the upper bound of the error between local stochastic gradient and the ground-truth one equals to the per-sample variance divided by the local mini-batch size, that is $\mathbb{E}\left[\left\|\nabla{F_i\left(\tilde{W}_i^t\right)}-G_i^{\rm t}\right\|^2\right]=\frac{\phi^2}{K_i^t}$.
To this end, we have:
\begin{equation}\tag{A-7}\label{A-7}
	\begin{aligned}
		\mathbb{E}\left[\|\epsilon^t\|^2\right]\le&\frac{3IU^2}{(K^t)^2}\sum_{i=1}^I(K_i^t)^2+\frac{3I\beta^2D^2}{(K^t)^2}\sum_{i=1}^I(K_i^t)^2\rho_i^t \\&+\frac{3I\phi^2}{(K^t)^2}\sum_{i=1}^IK_i^t.
	\end{aligned}
\end{equation}

With (\ref{A-5}) and (\ref{A-7}), by taking expectation at both sides of (\ref{A-3}), it holds that
\begin{equation}\label{A-8}\tag{A-8}
	\begin{aligned}
		&\mathbb{E}\left[F\left(W^{t+1}\right)\right]\\&\le F\left(W^t\right)\!-\frac{1}{2\beta}\!\left\|\nabla F\left(W^t\right)\right\|^2\!+\!\frac{3\beta ID^2}{(K^t)^2}\sum_{i=1}^I(K_i^t)^2\rho_i^t\\&\ \ \ +\!\frac{Q{a^{\prime}}^2\sigma^2\!+\!bIM^2}{\beta}\!+\!\frac{3IU^2}{\beta (K^t)^2}\sum_{i=1}^I(K_i^t)^2\!+\!\frac{3I\phi^2}{\beta (K^t)^2}\sum_{i=1}^IK_i^t.
	\end{aligned}
\end{equation}

Based on (\ref{A-8}) and $K^t=\sum_{i=1}^I K_i^t$, we can rewrite the upper bound of $\mathbb{E}\left[\|\!\nabla F\left(W^{t}\right)\!\|^2\right]$ as
\begin{equation}\label{A-9}\tag{A-9}
	\begin{aligned}
		&\frac{\mathbb{E}\left[\|\nabla F\left(W^{t}\right)\|^2\right]}{2\beta}\\&\le F\left(W^{t}\right)\!-\!\mathbb{E}\left[F\left(W^{t+1}\right)\right]\!+\!\!\frac{3\beta ID^2}{(K^t)^2}\sum_{i=1}^I(K_i^t)^2\rho_i^t\!\\&\ \ +\!\frac{Q{a^{\prime}}^2\sigma^2+bIM^2}{\beta}+\!\frac{3IU^2}{\beta (K^t)^2}\sum_{i=1}^I(K_i^t)^2+\frac{3I\phi^2}{\beta K^t}.
	\end{aligned}
\end{equation}
Summing up above inequalities from $t=0$ to $t=S$ and then averaging yields
\begin{equation}\label{A-10}\tag{A-10}
	\begin{aligned}
		&\frac{1}{S+1}\sum_{t=0}^S\mathbb{E}\left[\|\nabla{F\left({W}^t\right)}\|^2\right]\\&\le\! \frac{\sum_{t=0}^S\{\mathbb{E}\left[{F\left(W^t\right)}\right]\!-\!\mathbb{E}\left[F\left(W^{t+1}\right)\right]\}}{\left(S+1\right)/(2\beta)}\!+\!2(\!Q{a^{\prime}}^2\sigma^2\!+\!bIM^2\!)\\&\ \ +\mathbb{E}_t\left[\!\frac{6\beta^2 ID^2}{(K^t)^2}\sum_{i=1}^I(K_i^t)^2\rho_i^t+\frac{6IU^2}{ (K^t)^2}\sum_{i=1}^I(K_i^t)^2+\frac{6I\phi^2}{K^t}\right]\\
		&\le\frac{{F\left(W^0\right)}-F\left(W^*\right)}{\left(S+1\right)/(2\beta)}+2(Q{a^{\prime}}^2\sigma^2+bIM^2)\\&  \ \ +\mathbb{E}_t\!\left[\!\frac{6\beta^2 ID^2}{(K^t)^2}\sum_{i=1}^I(K_i^t)^2\rho_i^t\!+\!\frac{6IU^2}{ (K^t)^2}\sum_{i=1}^I(K_i^t)^2\!+\!\frac{6I\phi^2}{K^t}\right],
	\end{aligned}
\end{equation}
where $\mathbb{E}_t\left[\cdot\right]$ calculates the average value during $S+1$ rounds.
This completes the proof.

\bibliographystyle{IEEEtran}
\bibliography{IEEEabrv,ref}

\end{document}